\documentclass[draft, english]{volcanica-template} 
\newcommand{\Title}{RF-free  control of nuclear spins with color centers in silicon carbide} % Manuscript title goes here.
\newcommand{\shortTitle}{} % Short title for header goes here, less than 60 characters.

\newcommand{\Author}{R. W\"ornle et al.\xspace} % First author goes here.
\author[{{\affiliation{1},\affiliation{2},\affiliation{3}}}]
{Raphael W\"ornle\Email{raphael.woernle@pi3.uni-stuttgart.de}}
\author[{{\affiliation{1},\affiliation{2}}}]
{Jonathan K\"orber}
\author[{{\affiliation{1},\affiliation{2}}}]
{Timo Steidl}
\author[{{\affiliation{4}}}]
{Georgy V. Astakhov}
\author[{{\affiliation{1},\affiliation{2}}}]
{Durga B. R. Dasari}
\author[{{\affiliation{5},\affiliation{6}}}]
{Florian Kaiser}
\author[{{\affiliation{1},\affiliation{2}}}]
{Vadim Vorobyov}
\author[{{\affiliation{1},\affiliation{2},\affiliation{3}}}]
{J\"org Wrachtrup}
\affil[{{\affiliation{1}}}]{					% affiliation #1
3rd Institute of Physics, University of Stuttgart, Allmandring 13, 70569 Stuttgart, Germany.
}

\affil[{{\affiliation{2}}}]{					% affiliation #2
Center for Integrated Quantum Science and Technology, 70569 Stuttgart, Germany.
}

\affil[{{\affiliation{3}}}]{                    % affiliation #3
Max Planck Institute for Solid State Research, Heisenbergstraße 1, 70569 Stuttgart, Germany.}
\affil[{{\affiliation{4}}}]{					% affiliation #4
Helmholtz-Zentrum Dresden-Rossendorf, Institute of Ion Beam Physics and Materials Research, 01328 Dresden, Germany.}

\affil[{{\affiliation{5}}}]{					% affiliation #5
Quantum Materials, Luxembourg Institute of Science and Technology (LIST), 28 Avenue des Hauts Fourneaux, 4362 Belval, Luxembourg.}

\affil[{{\affiliation{6}}}]{					% affiliation #6
University of Luxembourg, 2 Avenue de l'Université, 4365 Belval, Luxembourg.}

\usepackage{blindtext}
\usepackage{braket}
\usepackage{upgreek}
\usepackage{comment}
\usepackage{ulem}
\usepackage{xcolor}
\begin{document}

%------------------------------------------------------------
%  Abstract(s) and Keywords
%------------------------------------------------------------	

\FrontMatter{\protect{
\noindent Color centers that enable nuclear-spin control  offer a powerful route towards simplified and scalable quantum devices. Such capabilities are especially valuable for quantum sensing and computing platforms that already find applications in biology, materials science, and geophysics. A key challenge is the coherent manipulation of nearby nuclear spins, which serve as quantum memories and auxiliary qubits but conventionally require additional high-power RF fields which increase the experimental complexity and overall power consumption. Finding systems where both electron and nuclear spins can be controlled using a single MW source is therefore highly desirable. Here, using a modified divacancy center in silicon carbide, we show that coherent control of a coupled nuclear spin is possible without any RF fields. Instead, MW pulses driving the electron spin also manipulate the nuclear spin through hyperfine-enhanced effects, activated by a precisely tilted external magnetic field. We demonstrate high-fidelity nuclear-spin control, achieving $89\%$ two-qubit tomography fidelity and nearly $T_1$-limited nuclear coherence times. This approach offers a simplified and scalable route for future quantum applications. 
}}[]{}
%-----------------------------------------------------------
%  Maintext
%------------------------------------------------------------	

\section*{INTRODUCTION}\label{sec:introduction}
4H-silicon carbide (SiC) has recently emerged as a promising platform to host point defects with possible applications in quantum technologies, such as distributed quantum computing or sensing \cite{Zhou2025SiliconNetworks, Castelletto2022SiliconTechnology}.

SiC contains a multitude of different color centers \cite{Son2020DevelopingSpintronics, Castelletto2020SiliconApplications}. One of the most studied color center is the silicon vacancy center (V$_{\mathrm{Si}}$), in particular the so-called V2 center is a promising candidate in the field of quantum communication due to its encouraging properties at low temperatures (LT) \cite{Babin2022FabricationCoherence,Morioka2020Spin-controlledCarbide, Simin2017LockingCarbide}. Other widely researched color centers are divacancies (V$_{\mathrm{Si}}$V$_{\mathrm{C}}$), the so-called PL1-4 centers (named after their four possible orientations within the 4H-SiC lattice) \cite{Koehl2011RoomCarbide, Christle2017IsolatedInterface, Falk2013PolytypeCarbide}. These defects have also achieved excellent results at cryogenic temperatures reaching seconds-long spin coherence times \cite{Anderson2022Five-secondCarbide, Christle2015IsolatedTimes}, as well as lifetime-limited optical linewidths \cite{Udvarhelyi2020Vibronic4H-SiC, Steidl2025SingleTemperature,Anderson2019ElectricalDevices}, both being essential for quantum communication and distributed quantum computing.
Current research investigates how to integrate these color centers into photonic (nano-)structures to maximize count rates, and thus application relevance \cite{Babin2022FabricationCoherence, Krumrein2024PreciseCarbide, Korber2024FluorescenceAntenna}.

Quite recently, a new family of color centers, modified divacancies in SiC, was discovered \cite{Li2022_PL6, Son2022Modified4H-SiC}. 
The current understanding of those modified divacancies is related to a nearby antisite which suggests that a reliable implantation and creation of those color centers is possible. This interpretation is supported by recent theoretical works proposing the correspondence of a divacancy coupled to a nearby carbon antisite at different crystallographic positions~\cite{Zhao2025AnalyzingScreening,Chen2025AtomicSpectroscopy}.

Out of those modified divacancy centers the so-called PL6 centers demonstrate with their high count rate from bulk emitters ($\sim$ 200 kcps) and high spin readout contrast ($\sim$ 30 \%) at room temperature \cite{Li2022_PL6}, that they can match the performance of the long-studied NV center in diamond \cite{Jelezko2006SingleReview, Doherty2013TheDiamond, Childress2006CoherentDiamond}, making them suitable candidates for potential room temperature sensing applications.
For PL6 centers, further improvements to 60 \% spin readout contrast already were demonstrated in highly strained SiC membranes \cite{Hu2025StrainMembranes}. Additionally, quantum technologies based on modified divacancies in SiC offer a pathway towards full CMOS compatibility and wafer-scale fabrication \cite{Weitzel1996SiliconDevices, Sharmila2025SiliconPerspectives}.
Besides this, modified divacancies such as the PL6 center are potentially well-suited for biology studies \cite{Li2025Non-invasiveQubits}, considering their emission spectrum in the second biological window (1000 - 1300 nm) due to reduced scattering and absorption in tissue \cite{Koehl2011RoomCarbide}.

\begin{figure*}[ht]
\includegraphics[width=\linewidth]{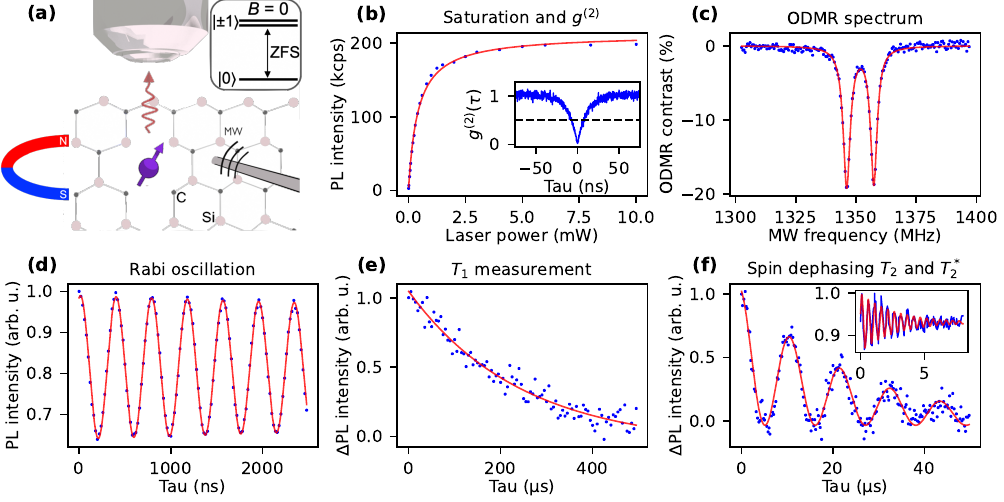}
\centering
\caption[]{\textbf{Properties of a single PL6 center.} \textbf{(a)} Schematic representation of a PL6 center with ground state level structure and the experimental apparatus with wire for MW, and external magnetic field and objective for excitation and collection of the fluorescence emission. \textbf{(b)} Saturation study of a single PL6 center with a saturation intensity of $211.7 \pm 1.8$ kcps. Inset shows the second order correlation function g$^{(2)}(\tau)$ confirming a clear single defect behavior. \textbf{(c)} CW-ODMR spectrum of a single PL6 center in zero magnetic field with a two-Lorentzian fit function centered around 1351.8 MHz.
\textbf{(d)} Rabi oscillation fitted with a damped cosine function. 
\textbf{(e)} Measurement of the spin-lattice relaxation time $T_1$ in a magnetic field of $B = 210 $ G. A single exponential fitting function was used to determine $T_1 = 242.8 \pm 22.1 \,\upmu$s. \textbf{(f)} Hahn echo measured in a magnetic field of $B = 210 $ G. $T_2$ is estimated from the damped cosine fitting to $T_2 = 25.0 \pm 1.3 \,\upmu$s. Inset shows Ramsey measurement measured at a detuning of 3 MHz yielding a pure spin-dephasing time $T_2^* = 2.7 \pm 0.3 \,\upmu$s.
% If figure twocolumn: Put * behind figure
}
\label{Fig1}
\end{figure*}

Recent progress has also been made on integrating PL6 centers into nanophotonic structures to further improve their count rates ~\cite{Hu2024Room-temperaturePlatform, Zhou2023Plasmonic-EnhancedMembranes, Bao2025DeterministicCavities}, and first cryogenic optical studies have been performed to probe their optical properties in more detail~\cite{He2024_LT}.

Due to their aforementioned favorable optical and spin properties, PL6 centers also enable sensing and control of nearby nuclear spins. Nuclear spins in the local environment couple to the electron spin of the defect, allowing them to be detected and used as qubit registers for quantum information storage. Conventionally, nuclear-spin manipulation requires strong radio-frequency (RF) pulses because of their weak magnetic moment. Eliminating RF fields in favor of weaker microwave (MW) pulses is therefore highly advantageous, reducing experimental complexity and power consumption, avoiding RF-induced heating, and minimizing thermal drift — particularly relevant in low-temperature measurements.

In this work, we demonstrate RF-free nuclear spin control enabled by precisely tilting the externally applied magnetic field, which activates hyperfine-mediated nuclear spin precession. The observed nuclear spin precession further provides a sensitive probe of the magnetic-field orientation, enabling accurate field alignment using the coupled nuclear spin itself.

\section*{RESULTS AND DISCUSSION}\label{sec:results}
\subsection*{Spin properties of PL6 centers}

Figure \ref*{Fig1}a illustrates a schematic representation of the measurement apparatus with a PL6 center.
A power-dependent saturation study on a single PL6 center, depicted in Figure~\ref{Fig1}b, yields a background-corrected saturation count rate of $I_\mathrm{S} = 211.7 \pm 1.8~\mathrm{kcps}$, consistent with previously reported values for unstructured bulk samples~\cite{Li2022_PL6}.
Autocorrelation measurements using a Hanbury-Brown and Twiss interferometer confirm with $g^{(2)}(0) \ll 0.5 $ that the signal originates from a single defect center, depicted in the inset in Figure \ref{Fig1}b and in Supplementary Figure S1.
In this work, individual PL6 centers are identified by their characteristic zero-field splitting (ZFS) measured through continuous-wave optically detected magnetic resonance (CW-ODMR) measurements. Figure~\ref{Fig1}c shows a typical CW-ODMR spectrum of a single PL6 center recorded at a laser power of 50~$\upmu$W in zero magnetic field, exhibiting its two characteristic ODMR dips. The data are fitted with a double Lorentzian function, yielding $f_1 = 1346.1~\mathrm{MHz}$ and $f_2 = 1357.4~\mathrm{MHz}$, corresponding to the $m\mathrm{_s} = 0 \rightarrow m\mathrm{_s} = \pm1$ transitions, with a measured contrast of approximately 19\,\%. This results in ZFS parameters $D= 1351.8 ~\mathrm{MHz}$ and $E= 5.6 ~\mathrm{MHz}$  corresponding to the Hamiltonian (see equation \ref{Ham1}), which are in good agreement with previous reports of PL6 centers~\cite{Li2022_PL6, Koehl2011RoomCarbide, Falk2013PolytypeCarbide, Falk2015OpticalCarbide}.

Using pulsed measurements, the spin readout contrast can be enhanced even more. Here, Rabi oscillation measurements show a contrast of 35\,\%, fitted with a damped cosine function, as shown in Figure~\ref{Fig1}d. 

The spin coherence times, critical for quantum technology applications, were characterized at an applied magnetic field of 210~G to suppress spin-bath decoherence. The spin-lattice relaxation time $T_1$ was determined to be $242.8 \pm 22.1~\upmu$s, while the spin-spin relaxation time $T_2$ was measured to be $25.0 \pm 1.3~\upmu$s, as shown in Figure~\ref{Fig1}e. The inhomogeneous spin-dephasing time $T_2^*$ is deduced to be $2.7 \pm 0.3~\upmu$s, displayed in the inset of Figure~\ref{Fig1}e. While the relaxation times $T_2$ and $T_2^*$ are comparable to previous reports, the measured $T_1$ is nearly twice as long as previously reported values at room temperature~\cite{Li2022_PL6, Hu2024Room-temperaturePlatform}. The improved relaxation times are attributed, in part, to the use of a higher intrinsic quality of our sample.

\subsection*{Coherent control of single nuclear spins}

\begin{figure}[ht]
\includegraphics[width=\linewidth]{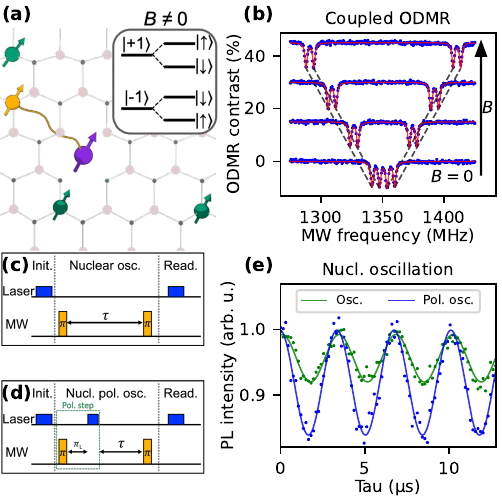}
\centering
\caption[]{\textbf{PL6 coupled to nearby nuclear spin.}
\textbf{(a)} Schematic representation of a single color center coupled to a nearby $^{13}$C nuclear spin and a coupling strength of 6.7 MHz with shown hyperfine splitting. \textbf{(b)} ODMR spectra of the coupled PL6 center spin without and with applied magnetic field up to a field strength of 20 G and corresponding Zeeman splitting. \textbf{(c)} Pulse sequence for nuclear oscillation. The $\ket{-1, \uparrow}$ transition from the ODMR measurement is used as a frequency for the MW pulse.
\textbf{(d)} Modified version of the pulse sequence shown in (c) with laser polarization step to further polarize the nuclear spin, shown first by Dutt et al. \cite{Dutt2007QuantumDiamond}. \textbf{(e)} Experimental results measured at an external magnetic field of 240 G for the shown pulse sequences without (\textit{green}) and with (\textit{blue}) polarized oscillation with corresponding fit functions.
% If figure twocolumn: Put * behind figure
}
\label{Fig2}
\end{figure}

Unlike diamond, SiC has the advantage of having a diatomic base consisting of silicon and carbon atoms. This means that it has more stable isotopes with a non-zero nuclear spin than diamond: $^{13}$C with a natural abundance of 1.1 \% and $^{29}$Si with 4.7 \%, respectively.
We found one PL6 center strongly coupled to a nearby $^{13}$C nuclear spin with a coupling strength of 6.7 MHz (Supplementary Figures S1-S3). 
A schematic representation of the coupling of the PL6 center with the nuclear spin is shown in Figure \ref*{Fig2}a.
The characteristic CW-ODMR splitting can be seen in Figure \ref*{Fig2}b without and with increasing external magnetic field. In contrast to an ODMR spectrum without coupled nuclear spins, the $m_{\mathrm{s}}= \pm 1 $ transitions are additionally split by the hyperfine interaction with the nuclear spin.
Inspired by Hu et al. \cite{Hu2024Room-temperaturePlatform}, we measured the nuclear spin precession using the sequence depicted in Figure \ref*{Fig2}c. \\

The experimental result is illustrated in green in Figure \ref*{Fig2}e. The precession can be observed without the need for an additional RF source.
First, we initialize the nuclear spin with a short laser pulse, after which we apply a spin-selective $\pi$ pulse on the electron spin (here, on the $\ket{-1, \uparrow}$) transition after which a waiting time as well as another $\pi$ pulse are applied before readout. %\sout{Quantitatively, the precession has a much slower oscillation compared to the electron Rabi oscillations. This is due to the nuclear spin precession being controlled only by the transversal external magnetic field whereas electron Rabi oscillation are driven by the applied microwave signal.}
{Quantitatively, the nuclear spin precession is much slower than the electron Rabi oscillations, primarily because the nuclear spin has a much smaller gyromagnetic ratio than the electron. Although both oscillation frequencies depend on magnetic field strength, they are governed by different fields: the nuclear precession frequency depends on the external magnetic field strength, while the electron Rabi frequency depends on the amplitude of the applied microwave field.}
Additionally, it becomes evident that the signal of the oscillation of the nuclear spin decays much slower than the Rabi measurement of the electron spin shown in Figure 1d.\\
This can be explained as follows: in contrast to manipulating electron spins, which is achieved through optical excitation and the application of microwave signals, manipulating nuclear spins additionally requires the ability to control the nuclear spin via the electron spin. When the electron spin is initialized by a laser pulse, the nuclear spin is in a mixed state of spin up and spin down. To polarize the nuclear spin, the nuclear spin is first mapped onto the electron spin by applying a microwave $\pi$ pulse, and then the electron spin is reinitialized by a short laser pulse, shown in Figure \ref*{Fig2}d, based on the previous work by Dutt et al. \cite{Dutt2007QuantumDiamond}. 
This results in the nuclear spin being polarized and the contrast being increased even further (Supplementary Figure S4). It is important that the laser polarization pulse after $T_{\text{Larmor}} = \omega_{\text{nucl}}/\pi$ is applied after the $\pi$ pulse to achieve maximum polarization. This is shown in blue in Figure \ref*{Fig2}e.
Since a similar control is achieved as in the previous work by Hu et al., but in a different measurement configuration without the application of an RF pulse, a more detailed investigation is required here.

\begin{figure*}[ht]
\includegraphics[width=\linewidth]{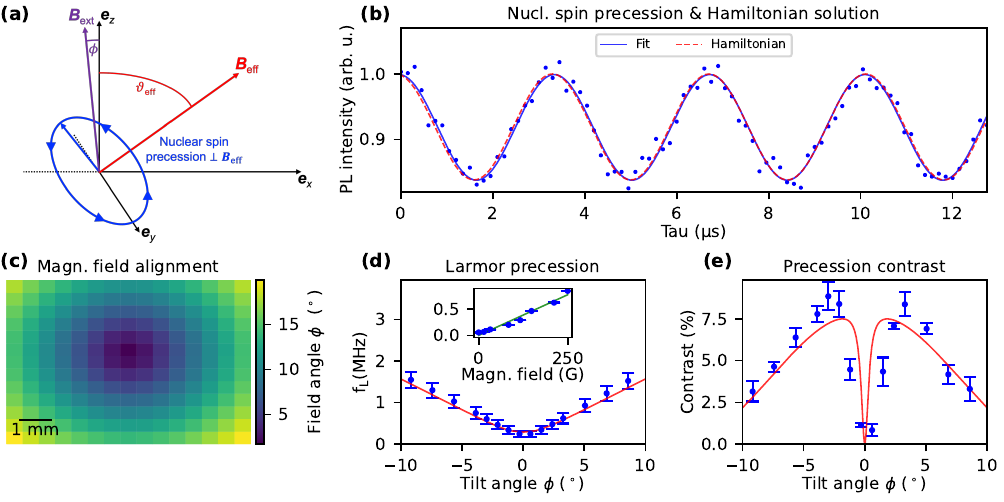}
\centering
\caption[]{\textbf{Theoretical derivation of nuclear oscillation.}
\textbf{(a)} Schematic representation of the tilt of the magnetic field causing the precision of the nuclear spin with enhanced oscillation amplitude due to gyromagnetic enhancement. \textbf{(b)} Nuclear oscillation experimental data (\textit{blue}) \& fit function (\textit{blue line}) with derived theoretical solution of the Hamiltonian (\textit{red dashed line}) at an external magnetic field of 240 G and a tilt angle $\phi =  2~^\circ$.
\textbf{(c)} Field angle $\phi$ calculated from experimental data for different planar alignments of the magnetic field. 
\textbf{(d)} Larmor frequency in dependence of the tilt angle $\phi$ of the magnetic field with theoretical data of the Hamiltonian. Inset shows the Larmor frequency of the nuclear oscillation in dependence of the external magnetic field strength with linear fit function $f_\mathrm{L} \propto B$.
\textbf{(e)} Contrast of the nuclear oscillation in dependence of the magnetic tilt angle $\phi$ in comparison to the theoretical values with a maximum contrast for a tilt angle $\phi = 1.8 ~ ^\circ$.
% If figure twocolumn: Put * behind figure
}
\label{Fig3}
\end{figure*}

\subsection*{Model}
The Hamiltonian of the system can, in general, be written as 
\begin{equation}\label{Ham1}
    H = D S_\mathrm{z}^2 
      + E \left( S_\mathrm{x}^2 - S_\mathrm{y}^2 \right)
      + \gamma_\mathrm{e}\, \mathbf{B} \cdot \mathbf{S}
      + \mathbf{S} \cdot \mathbf{A} \cdot \mathbf{I}
      - \gamma_{\mathrm{I}}\,  {\mathbf{B} \cdot \mathbf{I}}.
\end{equation}

where the experimentally determined axial zero-field splitting and the transverse (non-axial) component are  given by $D$ and $E$, respectively.
$\mathbf{S}$ and $\mathbf{I}$ denote the electron and nuclear spin operators, {$\gamma_\mathrm{e}$ and $\gamma_\mathrm{I}$ correspond to the gyromagnetic ratios of the electron spin and the $^{13}$C nuclear spin}. 
Additionally, $\mathbf{A}$ is the hyperfine interaction tensor and $\mathbf{B}$ is the 
externally applied magnetic field. The external magnetic field can be decomposed into components parallel and perpendicular to the quantization axis as shown in Figure 3a.

Analogous to Bürgler et al. \cite{Burgler2023All-opticalDiamond}, the effective Hamiltonian can now be determined using Van Vleck perturbation theory for 
the $m_\mathrm{s} = 0$ ground-state subspace 
$\{\ket{0,-1/2},\, \ket{0,+1/2}\}$ (The full derivation of the effective Hamiltonian is given in the Supplementary Information). 
In this treatment, the transverse component of the zero-field splitting is 
neglected as $E \ll D{, \gamma_\mathrm{e}B}$,  {that yields only a small energetic offset and mixing of the electron-spin manifolds which is suppressed by the longitudinal Zeeman splitting}.
The remaining corrections arise from virtual transitions to the 
$m_\mathrm{s} = \pm 1$ states induced by the transverse hyperfine interaction. 
This leads to the following effective Hamiltonian:
\begin{equation}\label{Hamiltonian}
    H_{\mathrm{eff}} = 
    \frac{1}{2}
    \begin{pmatrix}
        \gamma_\mathrm{I} B_\mathrm{z} + \nu_\mathrm{z} 
        &
        \gamma_\mathrm{I} B_\perp + \nu_\perp
        \\
        \gamma_\mathrm{I} B_\perp + \nu_\perp 
        &
        - \gamma_\mathrm{I} B_\mathrm{z} - \nu_\mathrm{z}
    \end{pmatrix},
\end{equation}
where the correction terms
\begin{align}\label{Correction}
    \nu_\mathrm{z} &= 
    \frac{\gamma_\mathrm{e} B_\mathrm{z}\, A_\perp^2}
         {D^2 - (\gamma_\mathrm{e} B_\mathrm{z})^2},
    \qquad
    \nu_\perp =
    -\, \frac{2 \gamma_\mathrm{e} B_\perp\, A_\perp\, D}
            {D^2 - (\gamma_\mathrm{e} B_\mathrm{z})^2}
\end{align}
represent second-order shifts of the diagonal and off-diagonal matrix elements, 
respectively. These terms originate from the mixing between the $m_\mathrm{s}=0$ and 
$m_\mathrm{s}=\pm1$ manifolds and therefore depend on the transverse hyperfine coupling 
$A_\perp$ as well as the tilt of the external magnetic field parameterized by 
$B_\perp = B \sin\phi$ (Figure 3a).

Diagonalizing equation~\ref{Hamiltonian} yields the nuclear precession frequency
\begin{equation}\label{Precession_frequency}
    f_{\mathrm{nucl}} 
    = \gamma_\mathrm{I}\, |\mathbf{B}_{\mathrm{eff}}|
    = \sqrt{
        (\gamma_\mathrm{I} B_\mathrm{z} + \nu_\mathrm{z})^2 
        + (\gamma_\mathrm{I} B_\perp + \nu_\perp)^2
    },
\end{equation}
as well as the tilt angle of the effective magnetic field,
\begin{equation}
    \vartheta_\mathrm{eff} 
    = \tan^{-1}
    \left(
        \frac{\gamma_\mathrm{I} B_\perp + \nu_\perp}
             {\gamma_\mathrm{I} B_\mathrm{z} + \nu_\mathrm{z}}
    \right).
\end{equation}

The transverse component of the magnetic field plays a central role in this 
dynamics: the term $\gamma_\mathrm{I} B_\perp$ drives nuclear mixing directly, 
while the electron-mediated correction $\nu_\perp$ enhances this effect for 
larger $A_\perp$ and reduces it for stronger applied magnetic fields. 
In contrast, the diagonal correction $\nu_\mathrm{z}$ modifies the effective nuclear 
Zeeman splitting. In the special case of a perfectly aligned magnetic field ($B_\perp = 0$) and a purely axial hyperfine tensor ($A_\perp = 0$), both correction terms vanish. The Hamiltonian becomes diagonal, and consequently no nuclear precession is 
observed.

Additionally, one can determine the contrast $C$ of the nuclear precession. 
While the ideal contrast of the nuclear spin oscillation is determined by the relative strength of the perpendicular ($\gamma_\mathrm{I} B_\perp + \nu_\perp$) and parallel ($\gamma_\mathrm{I} B_\mathrm{z} + \nu_\mathrm{z}$) fields, one must also account for the readout contrast of the electron spin, which degrades in the presence of a perpendicular field $B_\perp$. Using a simple two-level model for the electron, the final contrast of the nuclear spin oscillation can be written as {the product of the electron spin readout contrast and the nuclear precession amplitude}
\begin{equation}\label{Contrast}
    C =  C_\mathrm{e}(\phi) \circ C_\mathrm{n}(\phi) = C_\mathrm{e}(\phi) \circ\left(\frac{\gamma_\mathrm{I} B_\perp +\nu_\perp}{f_{\mathrm{nucl}}}\right)^2.
\end{equation}

The total observable nuclear spin contrast $C_\mathrm{n}$ is convoluted with the electronic readout contrast ($C_\mathrm{e}$) which decreases with increasing $B_\perp$. Within the two-level approximation this contrast can be simply written as $C_\mathrm{e}   \sim \left( \frac{\gamma_\mathrm{e} B_\mathrm{z}-A_\parallel}{f_{\mathrm{e}}}\right)^2$, where the detuned electron spin precession frequency $f_{\mathrm{e}} = \sqrt{(\gamma_\mathrm{e} B_\mathrm{z}-A_\parallel)^2+(\gamma_\mathrm{e} B_\perp)^2}$. \\
{Quantitatively, equation \ref{Contrast} can be simplified in terms of the effective tilt angle $\vartheta_{\mathrm{eff}}$ to
\begin{equation}\label{Contrast_simplified}
    C \sim \cos^2\theta_\mathrm{e} \, \sin^2\vartheta_{\mathrm{eff}},
\end{equation}
where $\theta_\mathrm{e}$ is the tilt of the effective electronic quantization axis. A more detailed derivation is given in Supplementary Information.
}

The relation between the different orientations of the external and effective magnetic field, the defect orientation and the nuclear spin precession is shown schematically in Figure \ref{Fig3}a. The external magnetic field is slightly tilted relative to the crystal axis. This tilt causes the effective magnetic field to be strongly tilted relative to the external magnetic field. This leads to the precession of the nuclear spin observed in Figure \ref{Fig2}e.
Substituting the experimental values for the measured precession of the nuclear spin in Figure \ref{Fig2}e yields the Hamiltonian solution, which is shown in Figure \ref{Fig3}b with the experimental results and the fit function. It can be seen that the theoretical values are in very good agreement with the experimental ones. In this case, a magnetic tilt of $\phi = 2 \,^\circ$ was determined for the defect with the hyperfine couplings of $A_\mathrm{z} = 6.7$ MHz and $A_\perp = 5.5$ MHz.\\

\subsection*{Magnetic field alignment}
Precise alignment of the external magnetic field relative to the defect axis is crucial for both electron and nuclear spin control. Several established methods exist, including alignment via ODMR transitions, analysis of the spin Hamiltonian, or monitoring the fluorescence count rate of the color center \cite{Balasubramanian2008NanoscaleConditions} (Supplementary Figure S5). The alignment of the field using the spin Hamiltonian is shown as an example in more detail in Figure \ref{Fig3}c.
In this work, we demonstrate that nuclear spin precession provides a complementary and sensitive method to determine the field orientation.

The nuclear spin precession frequency and the observable contrast of its oscillations are strongly dependent on the tilt angle $\phi$ of the external field relative to the defect axis. As derived from the effective Hamiltonian, the precession frequency is primarily determined by the longitudinal component of the effective field, $f_\mathrm{nucl} \sim \gamma_\mathrm{I} B_\mathrm{z} + \nu_\mathrm{z}$, with small quadratic corrections from the transverse component $\nu_\perp$. In contrast, the oscillation amplitude (contrast) scales linearly with the transverse component, $C_\mathrm{n} \propto (\gamma_\mathrm{I} B_\perp + \nu_\perp)^2$, while simultaneously being modulated by the electron spin readout, which decreases for increasing $B_\perp$ (see Eq.~\ref{Contrast}). Consequently, nuclear spin precession and electron readout contrast exhibit opposing dependencies on the perpendicular field component.

Experimentally, this interplay results in a “sweet spot” for magnetic field alignment: a small, finite tilt of the external field maximizes the nuclear spin contrast, while keeping the oscillation frequency moderate, as illustrated in Figures~\ref{Fig3}d and \ref{Fig3}e. At zero tilt, the nuclear spin sees only a longitudinal field and the precession contrast vanishes, even though the Larmor frequency $f_0$ remains finite. As the tilt increases, the precession amplitude grows linearly with $\phi$, but the electron readout contrast decreases due to misalignment of the electron spin eigenstates. By fitting the experimentally measured nuclear precession frequency and contrast to the theoretical model, we extract both the {optimum tilt angle $\phi =  1.8\,^\circ$} and the hyperfine couplings $A_\mathrm{z} = 6.7$~MHz and $A_\perp = 5.5$~MHz, demonstrating excellent agreement between theory and experiment.

For practical applications, a further consideration arises from the trade-off between oscillation speed and signal-to-noise ratio (SNR). Faster nuclear precession allows more operations to be performed within the nuclear coherence time $T_2^{*, \mathrm{Nucl}}$, which is advantageous for quantum control and sensing. However, faster precession typically requires a larger transverse effective field, which, as discussed, reduces the electron readout contrast, and thereby the observed nuclear oscillation amplitude and SNR. Consequently, the experimental protocol must balance a sufficiently fast precession for timely operations with a large enough contrast to ensure high-fidelity readout. The sweet spot for nuclear spin contrast naturally accounts for this trade-off, providing an optimal tilt that maximizes the observable oscillation amplitude without excessively compromising the electron spin readout.

\begin{figure*}[ht]
\includegraphics[width=\linewidth]{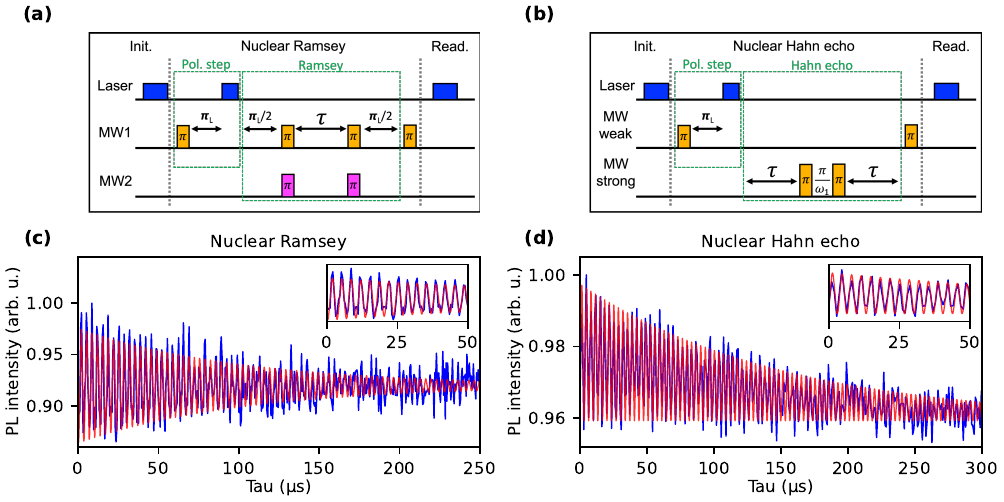}
\centering
\caption[]{\textbf{Nuclear Ramsey and spin echo.}
\textbf{(a) \& (b)} Schematic pulse sequence for the Ramsey ($T_2^{*, \mathrm{Nucl}}$) and nuclear spin echo ($T_2^{\text{Nucl}})$ measurement. For the Ramsey measurement, both microwave transitions $\ket{1,\uparrow}$ and $\ket{-1,\uparrow}$ are applied at the same time for driving the nuclear spin. For the nuclear spin echo measurement, two "fast" $\pi$ pulses are applied for Hahn echo, as first performed by Dutt et al. \cite{Dutt2007QuantumDiamond}.  \textbf{(c) \& (d)} Experimental result for the Ramsey and nuclear spin echo measurement yielding a spin dephasing time $T_2^{*, \mathrm{Nucl}} = 102.2 \pm 7.2 \,\upmu$s and a spin echo time for the nuclear spin of $T_2^{\mathrm{Nucl}} = 151.0 \pm 6.9 \,\upmu$s. Each inset shows a zoomed in version of the first 50 $\upmu$s. 
% If figure twocolumn: Put * behind figure
}
\label{Fig4}
\end{figure*}

\subsection*{Nuclear spin relaxation and coherence times}
We further demonstrate the application of the nuclear spin as a quantum memory. By transferring the electron-spin state onto the nuclear spin, the observed signal decays dramatically more slowly than the corresponding electron-spin coherence, highlighting a substantially longer nuclear coherence time.
The spin decoherence time $T_2^{^*,\text{Nucl}}$ is of particular interest here, as it is a well known measure to determine the magnetic sensitivity, an important quantitative criterion benchmark for potential applications in quantum sensing and metrology. The sensitivity can be determined quantitatively via the spin decoherence time to $    \eta \propto \nicefrac{1}{T_2^*}$ \cite{Barry2020SensitivityMagnetometry}.
Here, the pure dephasing time of the nuclear spin can be determined. Therefore, the sequence from Figure 2d gets modified in a way that both microwave transitions $\ket{1,\uparrow}$ and $\ket{-1,\uparrow}$ are applied at the same time for driving the nuclear spin. The corresponding sequence is illustrated in Figure \ref{Fig4}a. The measurement with a damped cosine fit function is depicted in Figure \ref{Fig4}c and results in a spin decoherence time $T_2^{^*, \text{Nucl}} = 102.2 \pm 7.2 \,\upmu$s. Compared to the electron spin, the nuclear spin has a T$_2^*$ that in this case is more than 30 times longer, a significant improvement for the magnetic sensitivity.

To access the nuclear-spin coherence, the pulse sequence in Figure~\ref{Fig4}a is modified by inserting two fast $\pi$ pulses on the selected ODMR transition, separated by a time interval corresponding to the hyperfine splitting, i.e.\ $\Delta t = \pi/\omega_{1}$ with $\omega_{1} = 2 \pi \cdot 6.7\, \mathrm{MHz}$. The resulting sequence is shown in Figure~4b. Experimentally, this produces a pronounced nuclear-spin echo whose decay time exceeds that of the electron spin. As shown in Figure~4d, the signal is well described by a damped squared-cosine fit, yielding a nuclear-spin relaxation time of $T_{2}^{\mathrm{Nucl}} = 151.0 \pm 6.9 \,\upmu$s, approximately six times longer than the electron-spin $T_{2}$. This extended coherence is consistent with the nuclear-spin dynamics being limited by the electron-spin $T_{1}$, an effect previously reported for NV centers in diamond \cite{Burgler2023All-opticalDiamond, Metsch2019InitializationDiamond, Broadway2018HighControl}.

\subsection*{Quantum state tomography (QST) under a tilted effective field}

Finally, we explore whether the method presented allows for a full set of quantum operations, sufficient, e.g., to perform quantum state tomography. Earlier, the nuclear spin Hamiltonian in the presence of a tilted magnetic field
always contains a non-zero longitudinal component, 
\(
H_{\mathrm{eff}} = \frac12(\Omega_\mathrm{z} \sigma_\mathrm{z} + \Omega_\mathrm{x} \sigma_\mathrm{x}),
\)
with $\Omega_\mathrm{z} = \gamma_\mathrm{I} B_\mathrm{z} + \nu_\mathrm{z}$ and 
$\Omega_\mathrm{x} = \gamma_\mathrm{I} B_\perp + \nu_\perp$.  
As a result, the Bloch sphere is not fully accessible: starting from 
$|0\rangle$, the spin states after nominal $\pi/2$ and $\pi$ pulses 
are now given by
\begin{align}
|\psi_{\pi/2}\rangle &= 
\frac{1}{\sqrt{2}}\Big[(1-\mathrm{i}\cos\vartheta_{\mathrm{eff}})|0\rangle - \mathrm{i} \sin\vartheta_{\mathrm{eff}} |1\rangle\Big],\\
|\psi_\pi\rangle &= -\mathrm{i}\Big[\cos\vartheta_{\mathrm{eff}}\,|0\rangle + \sin\vartheta_{\mathrm{eff}}\,|1\rangle\Big],
\end{align}
with $\vartheta_{\mathrm{eff}} = \tan^{-1}(\Omega_\mathrm{x}/\Omega_\mathrm{z})$.  
In such conditions, preparing ideal superposition states and performing ideal single qubit gates such as the Hadamard gate is challenging.

{Despite these limitations, state tomography remains possible by exploiting the fact that the effective Hamiltonian generates controlled rotations about a fixed tilted axis in the $XZ$-plane. Since direct projective readout is only available in the computational ($z$) basis, different measurement settings are realized by allowing the system to evolve for calibrated durations under $H_{\mathrm{eff}}$ prior to readout. This maps the computational-basis measurement onto effective projections along rotated directions in the $XZ$-plane.}

{In particular, evolution times $t = 0, ~\tau, ~2\tau$ yield three distinct, linearly independent measurement directions in the $XZ$-plane, viz., $\hat{\boldsymbol{n}}_1 = (0,0,1), ~
\hat{\boldsymbol{n}}_2 = (\sin \vartheta_{\mathrm{eff}},0,\cos \vartheta_{\mathrm{eff}}), ~
\hat{\boldsymbol{n}}_3 = (\sin 2\vartheta_{\mathrm{eff}},0,\cos 2\vartheta_{\mathrm{eff}})
$. These measurement directions provide sufficient linear independence to reconstruct the nuclear Bloch vector within the $XZ$-subspace, since the dynamics are confined to this plane in the absence of direct $\sigma_y$ control. While the use of three settings constitutes an overdetermined reconstruction scheme, it improves stability against experimental uncertainties in the effective rotation angle and measurement noise.}

{The $Y$-component of the Bloch vector is not directly accessible within this subspace but can be reconstructed by transferring the nuclear spin into the alternative electronic manifold, where the dynamics reduce to a pure phase evolution governed by the additional detuning induced by $A_\parallel$ (see Supplementary Information). Combining this phase evolution with the above measurement protocol enables full single-qubit state tomography. In the following, we demonstrate the protocol experimentally for selected cases and provide a detailed analysis in the Supplementary Information.}

\begin{figure}[ht]
\includegraphics[width=\linewidth]{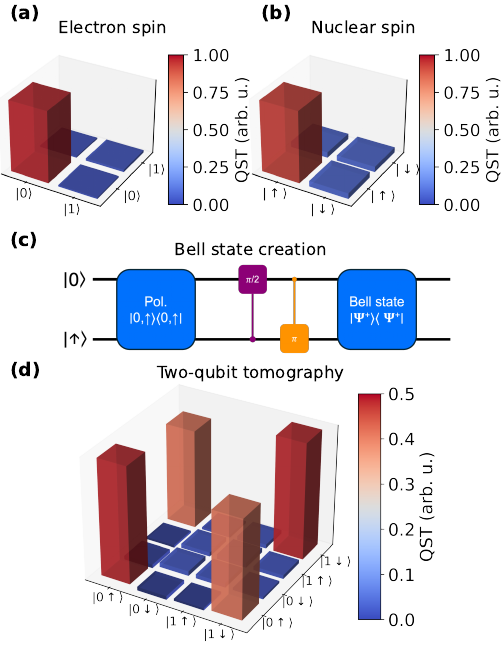}
\centering
\caption[]{\textbf{Quantum state tomography.}
Single qubit tomography for the electron spin \textbf{(a)} and the nuclear spin \textbf{(b)}  with reconstructed density matrix shown and a corresponding gate fidelity of 97 \% and 93 \%, respectively.
\textbf{(c)} Schematic representation of entanglement generation. A global $\nicefrac{\pi}{2}$ MW pulse on the electron spin and a subsequent $\pi$ pulse applied as a waiting time are used to create the Bell state from the polarized initial state.
\textbf{(d)} Two-qubit tomography for the nuclear spin coupled PL6 center yielding a fidelity of 89 \%. 
% If figure twocolumn: Put * behind figure
}
\label{Fig5}
\end{figure}

First, we show a one-qubit gate for the electron and nuclear spin individually, depicted in Figures \ref{Fig5}a and b, respectively.
From the corresponding reconstructed density matrices, a fidelity of 97 \% can be determined for the electron spin and 93 \% for the nuclear spin, showing high individual control over both single spins. Finally, a maximally entangled state, a Bell state $\ket{\Psi^+} = (\ket{0,\uparrow} + \ket{1, \downarrow})/\sqrt{2}$ is generated by applying two additional gates, shown in Figure \ref{Fig5}c.
The corresponding quantum operations are used as a combination of the MW pulses and waiting times already used in the one-qubit gates. In both cases, the $Z$-component can be read out straight forward without any additional sequences. For the electron spin manipulation, a MW phase shifter is used to apply $90~^\circ$ MW pulses in the $X$- and $Y$-plane. For the nuclear spin, the $X$-component can be manipulated by applying a waiting time corresponding to a $90~^\circ$ rotation. As mentioned before, the $Y$-component is accessed through applying two MW $\pi$ pulses to generate a phase onto the nuclear spin with a subsequent $90~^\circ$ rotation waiting time. By combining each three gate operations for the electron spin with the three nuclear spin gate operations, the two-qubit tomography can be performed. The resulting density matrix is shown in Figure \ref{Fig5}d. Here, also, a fidelity of 89 \% can be determined yielding a similar result to that of Hu et al. \cite{Hu2024Room-temperaturePlatform}, which shows a high coherent control of the entangled state through the driving of the nuclear spin but without any external RF-signal. The corresponding data as well as the detailed measurement sequences of the quantum state tomography are presented in the Supplementary Tables S1-S3 and Supplementary Figures S6-S8.

\subsection*{Summary and Outlook}
In conclusion, we have demonstrated coherent control of a nuclear spin in SiC at room temperature by utilizing a strongly coupled PL6 center without the necessity of a direct, RF-assisted driving of the nuclear spin. The high fidelity control of the nuclear spin is enabled by using its precession in a precisely tilted magnetic field, combined with the coherent control of the electron spin of a strongly coupled PL6 center, that is used for initialization and read out. Here, the high photon count rate and high spin readout contrast of the PL6 center at room temperature are key to our results. Additionally, state-of-the-art gate fidelities are achieved for a one- and two-qubit tomography reaching 97 \% for a single qubit gate and 89 \% for the two-qubit tomography. Furthermore, we showed that we can also use the nuclear spin as storage, thus achieving more extensive information storage than with the rapidly decohering electron spin of a single PL6 center. Here, a nearly $T_1$-limited nuclear spin coherence time of $T_{2}^{\mathrm{Nucl}} = 151.0 \pm 6.9 \,\upmu$s is achieved, being a factor of 6 higher compared to the electron spin coherence time. This method should also be easily extendable to the coupling of multiple nuclear spins, thus paving the way for the scalable development of defects in SiC for room temperature quantum applications.

%\input{Chapters/04-Bla}
%\textit{Vllt schteiben, dass der Kernspin nicht direkt ausgelesen werden kann und deshalb immer auf den Elektronenspin projeziert wird.}

\section*{Methods}
\paragraph*{Sample fabrication}
The sample of this study is a natural abundance SiC wafer with a 10 $\upmu$m-thick grown epi-layer (TySiC) with a doping concentration of [N] = $8.13\cdot 10^{15}$ cm$^{-3}$.
Nitrogen ions were implanted at 30 keV with a dose of $2.5 \cdot 10^{11} \, \nicefrac{\text{ions}}{\text{cm}^2}$ to create the PL6 color centers. Simulations show a depth of
defects of approximately 50 nm below the surface \cite{Ziegler2010SRIM2010}.
After the ion implantation, the sample was subjected to a two-step annealing process in a vacuum ($p<10^{-6}$ mbar), first for two hours at 500 $^{\circ}$C and then for one hour at 900 $^{\circ}$C.

\paragraph*{Optical characterization and spin manipulation}
All measurements were performed at room temperature.
For the measurements, a home-built confocal microscopy setup is used. A 940 nm laser (Thorlabs, M9-940-0200) was used for excitation, and the fluorescence was collected with single-photon detectors (SNSPDs, Photon Spot). The fluorescence emitted by the defects was filtered from the excitation light using a 1000 nm longpass dichroic filter and then further filtered through a 1000 nm longpass filter.
The excitation laser is pulsed with an acousto-optical modulator (AOM EQ Photonic 3200-124).
In all experiments, a 50 $\upmu$m thick copper wire is placed on the sample in close proximity to the implanted areas. 
The microwave signals are generated through vector signal generators (SMIQ03B \& SML03, Rohde \& Schwarz) and pass through multiple switches (ZASWA-2-50DR+, Mini Circuits) and a subsequent amplifier (ZHL-25W-272+, Mini Circuits) via SMA cables onto the wire attached to the sample. 90$^\circ$ phase shifters (ZX10Q-2-34-S+, Mini Circuits) are used for phase control of the spins.
A permanent neodymium magnet was used to generate an external magnetic field. This magnet was mounted on a stage with three linear translation stages (Thorlabs) to enable a precise and reproducible movement of the magnet in all three dimensions. Additionally, a single PL6 center was used to calculate the magnetic field strength using ODMR measurements. 

\section*{DATA AVAILABILITY}
The data supporting the presented findings are available at the following repository: \url{https://doi.org/10.18419/DARUS-5965}.
\section*{CODE AVAILABILITY}
The python-based measurement and evaluation codes used for this study are not publicly available due to their experimental complexity and insufficient documentation for external use, but are available from the corresponding author upon reasonable request.
\section*{COMPETING INTERESTS}
%The authors declare no conflict of interest.
J.W. is an Editorial Board Member of npj Quantum Information. J.W. was not involved in the journal’s review of, or decisions related to, this manuscript. The other authors do not have a competing interest.
\section*{ACKNOWLEDGMENTS}
We acknowledge fruitful discussions and experimental help from Erik Hesselmeier-Hüttmann, Rouven Maier, Marcel Krumrein, Adil Dogan, Rainer Stöhr and Pierre Kuna. Furthermore, we thank Arnold Weible from the Max Planck Institute for Intelligent Systems for helping with the first attempts of the annealing procedure.\\
This research was supported by the German Federal Ministry of Research, Technology and Space via the project QVOL2 (Grant agreement No. 03ZU2110GB) as well as the project QSi2V (Grant agreement No. 13N16756).
The project received funding from the European Union’s Horizon Europe research and innovation program through the SPINUS project  (Grant agreement No. 101135699). 
This research was supported by the European Commission via the project C-QuEnS (Grant Agreement No. 101135359).
G.V.A. acknowledges the project Quantum Sensing for Fundamental Physics (QS4Physics) from the Innovation pool of the research field Helmholtz Matter of the Helmholtz Association as well as the IBC facilities at the HZDR for support.
F.K. acknowledges support by the Luxembourg National Research Fund (FNR) via the PEARL chair "AQuaTSiC" under grant agreement 17792569 as well as the project "SiCqurTech" under grant agreement 18253399. F.K. additionally acknowledges the European Research Council for the project "Q-Chip" under grant agreement 101171067, the European Union's Horizon 2020 Research and Innovation Programme via the QuantERA project "SiCqurTech" under grant agreement 101017733, and the Horizon Europe Programme for the Flagship project "QIA" under grant agreement 101102140.

\section*{AUTHOR CONTRIBUTIONS}
The project was conceived by R.W. and J.W. and supervised by F.K., D.D., V.V. and J.W.
R.W. and J.K. prepared the samples and performed the annealing.
R.W and G.V.A designed the ion implantation.
R.W., J.K, T.S. and F.K. designed and constructed the optical setup. 
The optical measurements were conducted by R.W. and analyzed by R.W., D.D. and J.W. 
D.D. performed the  simulations.
The manuscript was written by R.W., J.K., D.D., V.V. and F.K.
All authors contributed to the manuscript.

\EndMatter
%\printbibitembibliography

\end{document}

\title{Supplementary Information to: \\ RF-free control of nuclear spins with color centers in silicon carbide}

\author{Raphael W\"ornle
\Email{raphael.woernle@pi3.uni-stuttgart.de}}
\thanks{E-Mail: raphael.woernle@pi3.uni-stuttgart.de}
\affiliation{% affiliation #1
3rd Institute of Physics, University of Stuttgart, Allmandring 13, 70569 Stuttgart, Germany.
}
\affiliation{% affiliation #2
Center for Integrated Quantum Science and Technology, 70569 Stuttgart, Germany.
}
\affiliation{% affiliation #3
Max Planck Institute for Solid State Research, Heisenbergstraße 1, 70569 Stuttgart, Germany.
}

\author{Jonathan K\"orber}
%\thanks{These authors contributed equally to this work}
%\email[]{jonathan.koerber@pi3.uni-stuttgart}
\affiliation{% affiliation #1
3rd Institute of Physics, University of Stuttgart, Allmandring 13, 70569 Stuttgart, Germany.
}
\affiliation{% affiliation #2
Center for Integrated Quantum Science and Technology, 70569 Stuttgart, Germany.
}

\author{Timo Steidl}
%\thanks{These authors contributed equally to this work}
%\email[]{jonathan.koerber@pi3.uni-stuttgart}
\affiliation{% affiliation #1
3rd Institute of Physics, University of Stuttgart, Allmandring 13, 70569 Stuttgart, Germany.
}
\affiliation{% affiliation #2
Center for Integrated Quantum Science and Technology, 70569 Stuttgart, Germany.
}

\author{Georgy V. Astakhov}
%\thanks{These authors contributed equally to this work}
%\email[]{jonathan.koerber@pi3.uni-stuttgart}
\affiliation{% affiliation #4
Helmholtz-Zentrum Dresden-Rossendorf, Institute of Ion Beam Physics and Materials Research, 01328 Dresden, Germany.
}

\author{Durga B. R. Dasari}
%\thanks{These authors contributed equally to this work}
%\email[]{jonathan.koerber@pi3.uni-stuttgart}
\affiliation{% affiliation #1
3rd Institute of Physics, University of Stuttgart, Allmandring 13, 70569 Stuttgart, Germany.
}
\affiliation{% affiliation #2
Center for Integrated Quantum Science and Technology, 70569 Stuttgart, Germany.
}

\author{Florian Kaiser}
%\thanks{These authors contributed equally to this work}
%\email[]{jonathan.koerber@pi3.uni-stuttgart}
\affiliation{% affiliation #5
Quantum Materials, Luxembourg Institute of Science and Technology (LIST), 28 Avenue des Hauts Fourneaux, 4362 Belval, Luxembourg.
}
\affiliation{% affiliation #6
University of Luxembourg, 2 Avenue de l'Université, 4365 Belval, Luxembourg.
}

\author{Vadim Vorobyov}
%\thanks{These authors contributed equally to this work}
%\email[]{jonathan.koerber@pi3.uni-stuttgart}
\affiliation{% affiliation #1
3rd Institute of Physics, University of Stuttgart, Allmandring 13, 70569 Stuttgart, Germany.
}
\affiliation{% affiliation #2
Center for Integrated Quantum Science and Technology, 70569 Stuttgart, Germany.
}

\author{J\"org Wrachtrup}
%\thanks{These authors contributed equally to this work}
\affiliation{% affiliation #1
3rd Institute of Physics, University of Stuttgart, Allmandring 13, 70569 Stuttgart, Germany.
}
\affiliation{% affiliation #2
Center for Integrated Quantum Science and Technology, 70569 Stuttgart, Germany.
}
\affiliation{% affiliation #3
Max Planck Institute for Solid State Research, Heisenbergstraße 1, 70569 Stuttgart, Germany.
}

\date{\today} 
\maketitle

%\input{Sup-texts/Methods}
%\clearpage
\section{Single defect verification}
For the defects investigated in the main part, autocorrelation measurements were performed in the Hanbury-Brown and Twiss interferometry experiment. %The setup from Figure \ref{Fig_S1_Setup} was slightly modified by installing a custom-made fiber beam splitter after coupling the emitted fluorescence into a fiber and distributing the signal onto two detectors to determine the time correlation between two subsequent emitted photons of the defect. 
The background-corrected measurements are shown for the PL6 center in Supplementary Figure \ref{Fig_S2_g2}a and for the PL6 center with coupled nuclear spin in Supplementary Figure \ref{Fig_S2_g2}b. Assuming a three level system for our defect \cite{Zhou2023Plasmonic-EnhancedMembranes}, the theoretical second-order correlation measurement can be fitted using
\begin{equation}
    g^{(2)}(\tau) = \frac1N\Big(1-(1+a)\cdot\mathrm{e}^{-\frac\tau{\tau_1}}+a\cdot \mathrm{e}^{-\frac\tau{\tau_2}}\Big) + \frac{N-1}{N} \quad \text{with} \quad g^{(2)}(\tau) \stackrel{\tau \rightarrow 0}{=} 1 - \frac{1}{N},
\end{equation}
where $a, \tau_1$ and $\tau_2$ are the amplitude and the (anti-)bunching time constants and $N$ the number of emitters \cite{Kurtsiefer2000StablePhotons}.
The results in both cases yield values of $g^{(2)}(0) \ll 0.5$, which clearly confirms that each defect is a single emitter.

\begin{figure}[ht]
\includegraphics[width=0.8\linewidth]{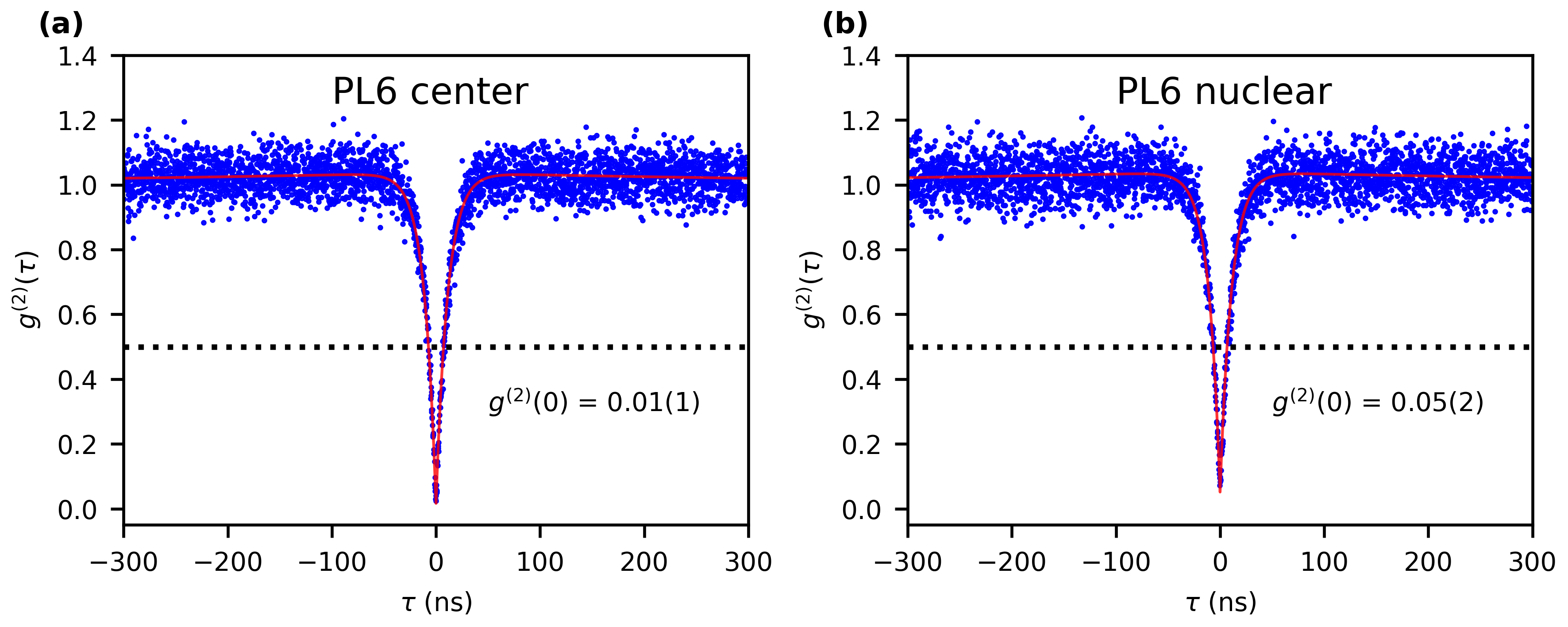}
\centering
\caption[]{\textbf{Autocorrelation measurements of PL6 centers.} Background-corrected autocorrelation measurements measured at an excitation laser power of 100 $\upmu$W for \textbf{(a)} a single PL6 center and \textbf{(b)} for the PL6 center coupled with a nearby nuclear spin. The fitted values for zero time delay yield a value of $g^{(2)}(0) = 0.01(1)$ for the PL6 center and $g^{(2)}(0) = 0.05(2)$ for the nuclear spin coupled PL6 center, both confirming the nature as single defects.
% If figure twocolumn: Put * behind figure
}
\label{Fig_S2_g2}
\end{figure}
\clearpage
\section{Nuclear spin control}
\subsection{Nuclear spin identification}

The type of the nuclear spin shown in Figure 2b in the main text can be identified experimentally. Hahn echo measurements are used for this purpose in a magnetic field of 145 G experimentally, which is illustrated in Supplementary Figure \ref{Fig_S2_Hahn}. The relationship between the Larmor frequencies $f_\mathrm{L} \propto \gamma_\mathrm{N} \cdot B$ results in a gyromagnetic ratio of $\gamma_\mathrm{N} = 10.88 \pm 0.12$ MHz/T from the measurement, which is close to the gyromagnetic ratio of $\gamma_\mathrm{N}^{^{13}\mathrm{C}} = 10.71$ MHz/T, therefore confirming the presence of a $^{13}$C nuclear spin.\\

\begin{figure}[ht]
\includegraphics[width=0.5\linewidth]{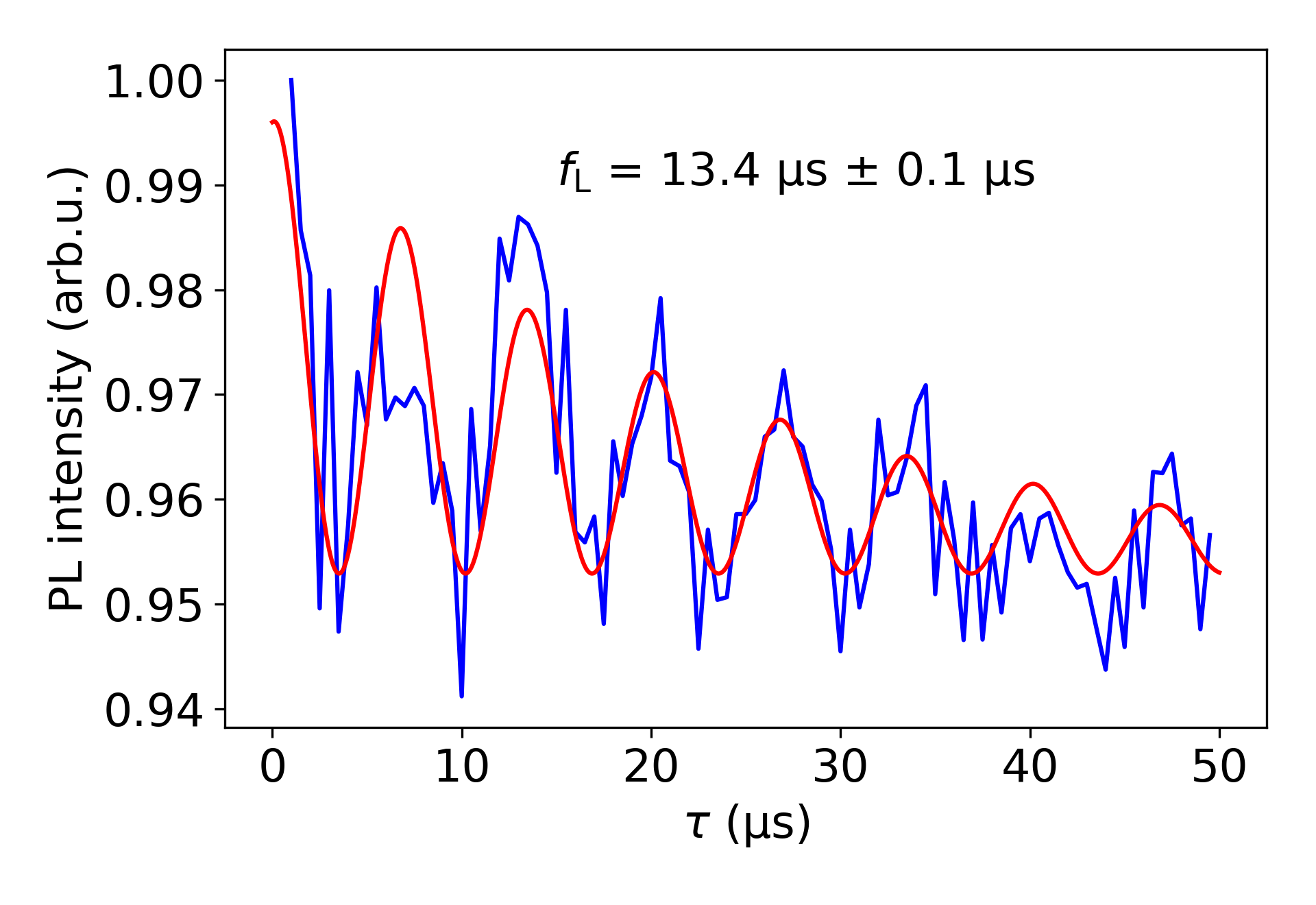}
\centering
\caption[]{\textbf{Nuclear spin identification.} Hahn echo measurement on the PL6 center with coupled nuclear spin in an external magnetic field of 145 G yielding a gyromagnetic ratio of $\gamma_\mathrm{N} = 10.88 \pm 0.12$ MHz/T matching the one for a $^{13}$C nuclear spin.
% If figure twocolumn: Put * behind figure
}
\label{Fig_S2_Hahn}
\end{figure}
%This sequence can now be modified, resulting in the Hahn echo measurements of the nuclear spin in the main part in Figure 4.

\subsection{Coupling strength}

Exact control over the nuclear spin can only be achieved if the phase between the two nuclear spin states can also be manipulated.
The exact strength of the coupling of the nuclear spin shown in Figure 2b in the main text can be determined using the following pulse sequence, shown in Supplementary Figure \ref*{Fig_S3_Nucl_Spin}a. The idea and implementation were taken from \cite{Childress2007CoherentState}.
For this purpose, the nuclear spin is initialized similarly as for the precession measurements.

\begin{figure}[ht]
\includegraphics[width=0.8\linewidth]{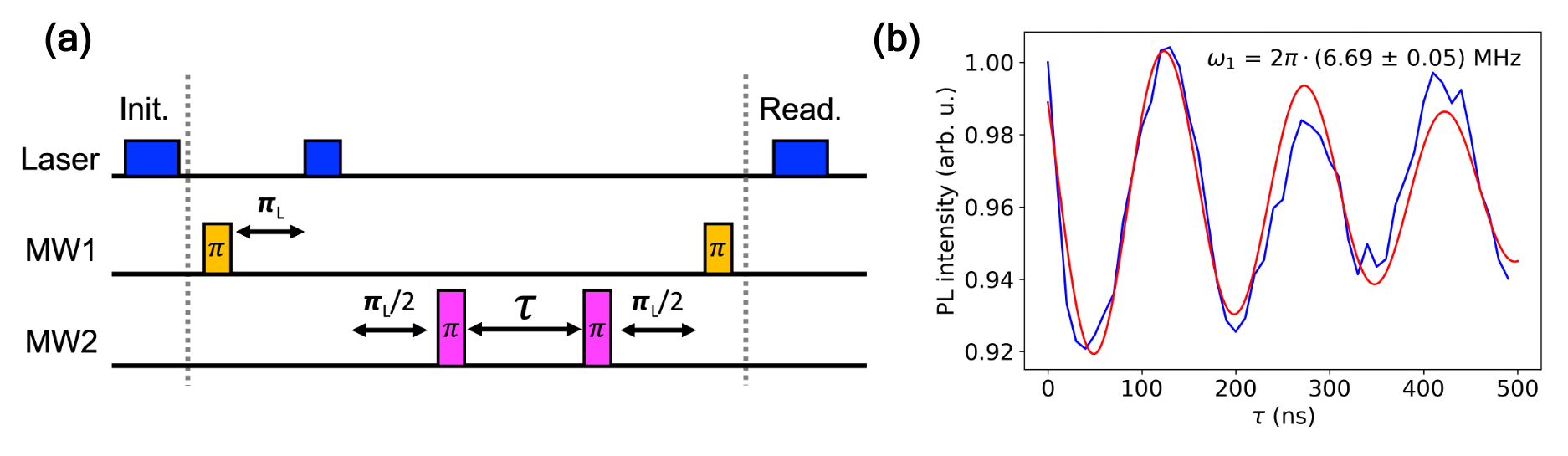}
\centering
\caption[]{\textbf{Nuclear spin coupling.} \textbf{(a)} Measurement sequence for determining the exact coupling strength of the coupled nuclear spin. Sequence adapted from \cite{Childress2007CoherentState}. \textbf{(b)} Experimental result including damped cosine fit yielding a coupling strength of $\omega_1 = 2 \pi \cdot 6.69(5) \,\mathrm{MHz}$.
% If figure twocolumn: Put * behind figure
}
\label{Fig_S3_Nucl_Spin}
\end{figure}
Furthermore, a strong, non-selective $\pi$ pulse excites both nuclear spin states; the subsequent waiting period before the second $\pi$ pulse creates a relative phase on the nuclear spin state. Subsequently, a weak $\pi$ pulse on the electron spin maps the system back onto the electron spin for read out, and the oscillations at the hyperfine coupling frequency $\omega_1$ can be observed. In this case, a coupling strength of $\omega_1 = 2 \pi \cdot 6.69(5) \,\mathrm{MHz}$ is determined and shown in Supplementary Figure \ref*{Fig_S3_Nucl_Spin}b.

This sequence can now be modified, resulting in the Hahn echo measurements of the nuclear spin in the main part in Figure 4.
\newpage

\subsection{Polarization}
The sequence with the polarization step shown in Figure 2c in the main part can be extended by repeating the polarization step $N$ times. The resonant driving of the nuclear spin polarizes it, thus improving the contrast. This is illustrated in Supplementary Figure \ref{Fig_S3_Polarization}a in dependence of the number of polarization steps.  It shows a saturation behavior whereby the contrast reaches almost twice the contrast of the CW-ODMR measurement seen in the main part in Figure 2b.

\begin{figure}[ht]
\includegraphics[width=\linewidth]{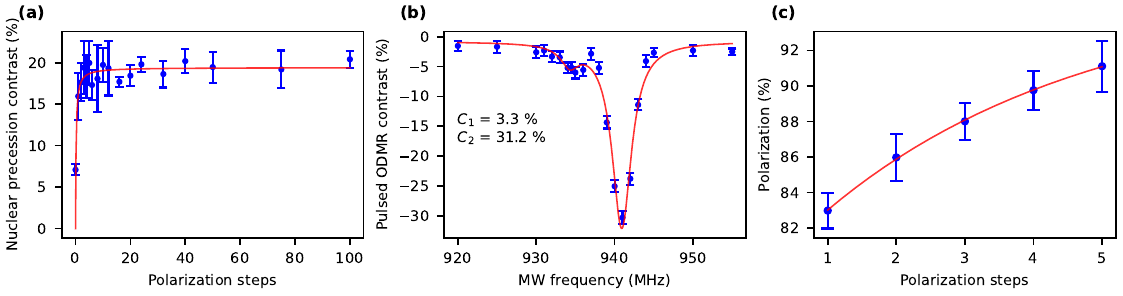}
\centering
\caption[]{\textbf{Polarization saturation study.} \textbf{(a)} Saturation study of the measured contrast of the nuclear precession in dependence of the number of polarization steps with saturation fit $C = \nicefrac{(C_{\mathrm{Sat}} \cdot N)}{(N+N_{\mathrm{Sat}})}$ yielding a saturation contrast of $19.4 \pm 0.7 \,\%$. Error bars are determined by the fitted contrast and fit error of each measurement.
\textbf{(b)} Pulsed ODMR measurement after previous polarization of nuclear spin yielding to a almost vanishing transition of the $\ket{-1, \uparrow}$ transition whereas the $\ket{-1, \downarrow}$ transition contrast is maximal.
\textbf{(c)} Polarization study of ratios of the pulsed ODMR contrasts in dependence of the number of polarization steps with corresponding fit function yielding a saturation polarization behavior of $95.5 \pm 0.8$ \%.

% If figure twocolumn: Put * behind figure
}
\label{Fig_S3_Polarization}
\end{figure}

Additionally, high polarization can be achieved through repeating the shown polarization step in Figure 2d in the main part multiple times, which has been shown to result in almost complete transition suppression. This is clearly achieved by performing $N = 5$ polarization steps and measuring pulsed ODMR. The result can be seen in Supplementary Figure \ref*{Fig_S3_Polarization}b, where a high degree of polarization is shown. In addition, a polarization study can be performed to determine the dependence of the polarization steps, as shown in Supplementary Figure \ref*{Fig_S3_Polarization}c. For this purpose, the polarization was determined from the ratio of the contrast of the two ODMR transitions. This results in a saturation polarization value of  approximately 95.5\% between the two states (in this case, $\ket{-1,\uparrow}$ and $\ket{-1,\downarrow}$). From this ratio, the effective zero-spin temperature can be determined using
\begin{equation}
    p_\uparrow = p_\downarrow \cdot \exp\Big(\frac{-\gamma_0 B}{k_\mathrm{B}T}\Big)
\end{equation}
with $\omega_0 = \gamma_0 B$ as the Larmor precession frequency. In our case, the effective nuclear spin temperature gets calculated to $T_{\text{nucl}} \approx 2.4 \,\mathrm{\upmu K}$. The idea was first shown by Dutt et al. (Suppl. S.2.4.2) \cite{Dutt2007QuantumDiamond}.

\clearpage
\section{Second-Order Perturbation Derivation}

In this section we derive the effective Hamiltonian acting on the
nuclear spin within the $m_\mathrm{s} = 0$ manifold of an $S=1$ electronic
spin coupled to a nuclear spin $I=1/2$.  
The derivation follows the Van Vleck (Schrieffer--Wolff) perturbation
procedure to second order, with all intermediate steps shown.

\subsection*{1. Full Hamiltonian and Hilbert-Space Partition}

The total spin Hamiltonian is given by
\begin{equation}
H = D S_\mathrm{z}^2 + E(S_\mathrm{x}^2 - S_\mathrm{y}^2)
+ \gamma_\mathrm{e} \mathbf{B}\cdot \mathbf{S}
+ \mathbf{S}\cdot \mathbf{A}\cdot \mathbf{I}
- \gamma_\mathrm{I} \mathbf{B}\cdot \mathbf{I},
\end{equation}
where $D \gg \gamma_\mathrm{e} B,\,A_\perp$.  
Here, $D$ and $E$ represent the experimentally determined axial zero-field splitting and the transverse (non-axial) component, 
$\mathbf{S}$ and $\mathbf{I}$ denote the electron and nuclear spin operators, $\gamma_\mathrm{e}$ and $\gamma_\mathrm{I}$ correspond to the gyromagnetic ratios of the electron spin and the $^{13}$C nuclear spin. Further, $\mathbf{A}$ is the hyperfine interaction tensor and $\mathbf{B}$ is the 
externally applied magnetic field.\\
In the following, as $E \ll D$, we neglect the transversal off-axial component of the zero-field splitting. This can be shown by a comparison of the PL6 system to the NV center in diamond. For the NV center, a $C_{3\mathrm{v}}$ symmetry holds. Therefore, the hyperfine tensor is axially symmetric and consists only out of the parallel and transversal parameters $A_{||}$ and $A_\perp$. Here, off-diagonal elements like $A_{\mathrm{zx}}$ vanish.\\
For the PL6 centers, things change slightly. Those color centers show $C_{1\mathrm{h}}$ symmetry with a slight tilt of 1-1.5$^\circ$. This makes them appear as axial color centers but with a small deviation from axial symmetry \cite{Son2022Modified4H-SiC}. As PL6 centers are almost axial, off-diagonal elements are neglected in this model.\\

We project the dynamics onto the two-dimensional subspace
\begin{equation}
\mathcal{H}_0 = \big\{ |0,\!\uparrow\rangle, |0,\!\downarrow\rangle \big\},
\end{equation}
which corresponds to the electronic ground state combined with the two nuclear spin orientations, while the electronic levels $|m_\mathrm{s}=\pm1\rangle$ form the excited subspace
$\mathcal{H}_1$.

We decompose the Hamiltonian as
\begin{equation}
H = H_0 + V ,
\end{equation}
where $H_0$ is chosen diagonal in the electronic ground state basis,
\begin{equation}
H_0 = D S_\mathrm{z}^2 + \gamma_\mathrm{e} B_\mathrm{z} S_\mathrm{z},
\end{equation}
and $V$ contains all transverse and coupling terms,
\begin{equation}
V =
\gamma_\mathrm{e}(B_\mathrm{x} S_\mathrm{x} + B_\mathrm{y} S_\mathrm{y})
+ \mathbf{S}\cdot\mathbf{A}\cdot\mathbf{I}
- \gamma_\mathrm{I} \mathbf{B}\cdot\mathbf{I}.
\end{equation}

Thus, the unperturbed eigenenergies are given by
\begin{align}
E_0 &= 0,\\
E_+ &= D + \gamma_\mathrm{e} B_\mathrm{z},\\
E_- &= D - \gamma_\mathrm{e} B_\mathrm{z}, 
\end{align}
depending on the externally applied magnetic field.
\subsection*{2. Matrix Elements of $S_\mathrm{x}$ and $S_\mathrm{y}$}

For an $S=1$ spin, one can calculate the expectation values for the $m_\mathrm{s}=0 \rightarrow m_\mathrm{s}=\pm1$ transitions  to 
\begin{align}
\langle 0|S_\mathrm{x}|\pm1\rangle &= \frac{1}{\sqrt{2}},\\
\langle 0|S_\mathrm{y}|\pm1\rangle &= \pm\frac{\mathrm{i}}{\sqrt{2}}.
\end{align}
Using the basis transformation 
\begin{equation}
B_\pm = B_\mathrm{x} \pm \mathrm{i} B_\mathrm{y}, 
\end{equation}
the transverse Zeeman interaction is determined by  
\begin{equation}
Z_\pm \equiv 
\langle 0| \gamma_\mathrm{e}(B_\mathrm{x} S_\mathrm{x} + B_\mathrm{y} S_\mathrm{y}) |\pm1\rangle
= \frac{\gamma_\mathrm{e}}{\sqrt{2}} B_\pm,
\end{equation}
depending on the transverse magnetic field.
\subsection*{3. Off-Axis Hyperfine Terms}
The hyperfine interaction tensor is decomposed into longitudinal and transverse components,
\begin{equation}
\mathbf{S}\cdot\mathbf{A}\cdot\mathbf{I}
= A_\parallel S_\mathrm{z} I_\mathrm{z} + A_\perp (S_\mathrm{x} I_\mathrm{x} + S_\mathrm{y} I_\mathrm{y}).
\end{equation}
Since $S_z|0\rangle=0$, the term with $A_\parallel$ vanishes in this subspace.
Thus, we get the hyperfine perturbation $V_{\text{hf}}$
\begin{equation}
V_{\text{hf},\perp} = A_\perp (S_\mathrm{x} I_\mathrm{x} + S_\mathrm{y} I_\mathrm{y})
= \frac{A_\perp}{2}(S_+ I_- + S_- I_+).
\end{equation}

Using the results from subsection 2, $\langle 0|S_\pm|\pm1\rangle = 1/\sqrt{2}$, we get the following results for the expectation values of the hyperfine perturbation:
\begin{align}
H_+ &= \langle 0|V_{\text{hf},\perp}|+1\rangle
= \frac{A_\perp}{\sqrt{2}} I_-,\\
H_- &= \langle 0|V_{\text{hf},\perp}|-1\rangle
= \frac{A_\perp}{\sqrt{2}} I_+.
\end{align}
\subsection*{4. Van Vleck / Schrieffer--Wolff Formula}

Let \(P\) denote the projector onto $\mathcal{H}_0$.
The effective Hamiltonian up to second order is thus given by:
\begin{equation}
H_{\rm eff} = 
P V P -
\sum_{\mathrm{m}=\pm1}
P V |\mathrm{m}\rangle\frac{1}{E_\mathrm{m} - E_0}\langle \mathrm{m}| V P.
\end{equation}

The first-order term gives the nuclear Zeeman interaction.
The second-order sum contains the hyperfine-enhanced corrections.

\subsection*{5. Diagonal Second-Order Correction ($\nu_z$)}
The diagonal second-order correction can be calculated using 
\begin{equation}
H_\pm H_\pm^\dagger
= \frac{A_\perp^2}{2}(I_\mathrm{x}^2 + I_\mathrm{y}^2)
= \frac{A_\perp^2}{2}\cdot \frac14,
\end{equation}

due to the spin identity of the carbon nuclear spin $I=1/2$.

Thus, $\nu_\mathrm{z}$ can be expressed by 
\begin{equation}
\nu_\mathrm{z} =
\frac{A_\perp^2}{2}
\left(\frac{1}{E_+}+\frac{1}{E_-}\right)
= A_\perp^2
\frac{E_+ + E_-}{2 E_+E_-}.
\end{equation}

Using the unperturbed eigenenergies,
\begin{equation}
E_+ + E_- = 2D, \quad
E_+ E_- = D^2 - (\gamma_\mathrm{e} B_\mathrm{z})^2,
\end{equation}
we can express the in the main text given second-order shift of the diagonal matrix element
\begin{equation}
\boxed{
\nu_\mathrm{z} = 
\frac{\gamma_\mathrm{e} B_\mathrm{z} A_\perp^2}
{D^2 - (\gamma_\mathrm{e} B_\mathrm{z})^2}.
}
\end{equation}

\subsection*{6. Off-Diagonal Second-Order Term ($\nu_\perp$)}
Off-diagonal contributions arise from mixed processes involving one transverse Zeeman interaction and one hyperfine interaction:
\begin{equation}
H_{\rm \mathrm{eff}}^{(2)} \supset
-\sum_{\mathrm{m}=\pm1} 
\frac{1}{E_\mathrm{m}}(Z_\mathrm{m} H_\mathrm{m}^\dagger + H_\mathrm{m} Z_\mathrm{m}^\dagger).
\end{equation}
In this case, both, the $\mathrm{m}=+1$ and $\mathrm{m}=-1$ contributions have to be considered using the identities and calculated matrix elements in the previous subsections. 

\begin{enumerate}
    \item \text{For $\mathrm{m}=+1$, we get the following contribution:}
\begin{equation}
Z_+ H_+^\dagger = \frac{\gamma_\mathrm{e} A_\perp}{2} B_+ I_+,\qquad
H_+ Z_+^\dagger = \frac{\gamma_\mathrm{e} A_\perp}{2} B_- I_-.
\end{equation}

\item \text{For $\mathrm{m}=-1$, we get the following contribution:}
\begin{equation}
Z_- H_-^\dagger = \frac{\gamma_\mathrm{e} A_\perp}{2} B_- I_-,\qquad
H_- Z_-^\dagger = \frac{\gamma_\mathrm{e} A_\perp}{2} B_+ I_+.
\end{equation}
\end{enumerate}

Summing both now yields:
\begin{equation}
\frac{\gamma_\mathrm{e} A_\perp}{2}
(B_+ I_+ + B_- I_-)
\left(\frac{1}{E_+}+\frac{1}{E_-}\right).
\end{equation}

Using the following identity:
\begin{equation}
B_+ I_+ + B_- I_- = 2(B_\mathrm{x} I_\mathrm{x} + B_\mathrm{y} I_\mathrm{y})
= 2 B_\perp I_\mathrm{x}
\quad (\text{choosing } \mathrm{x} \parallel B_\perp).
\end{equation}

we can put together the off-diagonal second-order correction term
\begin{equation}
\nu_\perp =
-\frac{2 \gamma_\mathrm{e} B_\perp A_\perp D}
{D^2 - (\gamma_\mathrm{e} B_\mathrm{z})^2},
\end{equation}
as given in the main text of the manuscript.
\subsection*{7. Final Effective Hamiltonian}

Putting all terms together yields now:
\begin{equation}
\boxed{
H_{\mathrm{eff}}
=
\frac12
\begin{pmatrix}
\gamma_\mathrm{I} B_\mathrm{z} + \nu_\mathrm{z} &
\gamma_\mathrm{I} B_\perp + \nu_\perp \\
\gamma_\mathrm{I} B_\perp + \nu_\perp &
-(\gamma_\mathrm{I} B_\mathrm{z} + \nu_\mathrm{z})
\end{pmatrix}.
}
\end{equation}

This is the effective nuclear-spin Hamiltonian used in the main text.
It contains both hyperfine-enhanced longitudinal energy shifts and
hyperfine-mediated transverse couplings arising from virtual
electronic excitations.
\clearpage
\section{General Hyperfine Tensor and Role of Off-Diagonal Components}

The hyperfine interaction for an arbitrary nearby $^{13}$C nucleus is

\begin{equation}
H_{\rm HF}
=
\sum_{i,j=x,y,z} S_i A_{ij} I_j .
\end{equation}

Unlike the intrinsic nitrogen spin in the NV center case \cite{Burgler2023All-opticalDiamond}, the hyperfine tensor of a
generic $^{13}$C nucleus is not required by symmetry to be diagonal in the
zero-field-splitting (ZFS) basis. We therefore consider the contribution of
all tensor components within the same Schrieffer--Wolff formalism, and show how they do not contribute to the observed effects.

\subsection*{Terms $A_{\rm zx}$ and $A_{\rm zy}$}

These terms are

\begin{equation}
V_{\rm zx}=A_{\rm zx}S_{\rm z}I_{\rm x},
\qquad
V_{\rm zy}=A_{\rm zy}S_{\rm z}I_{\rm y} .
\end{equation}

Since

\begin{equation}
\langle0|S_{\rm z}|\pm1\rangle=0 ,
\end{equation}

we have

\begin{equation}
PV_{\rm zx}Q=PV_{\rm zy}Q=0 .
\end{equation}

Therefore, these terms cannot participate in the virtual electronic
transitions required for the second-order Schrieffer--Wolff correction:

\begin{equation}
PV_{\rm zx}Q
\frac{1}{E_{\rm m}}
QV_{\rm zx}P
=
0 ,
\end{equation}

and similarly for $A_{\rm zy}$.

Hence,

\begin{equation}
\boxed{
A_{\rm zx},A_{\rm zy}
\text{ do not contribute to the effective Hamiltonian in the }m_{\rm s}=0
\text{ manifold at second order.}
}
\end{equation}

\subsection*{Terms $A_{\rm xz}$ and $A_{\rm yz}$}

The terms

\begin{equation}
V_{\rm xz}=A_{\rm xz}S_{\rm x}I_{\rm z} ,
\qquad
V_{\rm yz}=A_{\rm yz}S_{\rm y}I_{\rm z}
\end{equation}

connect the $m_\mathrm{s}=0$ and $m_\mathrm{s}=\pm1$ manifolds.

For example,

\begin{equation}
\langle0|V_{\rm xz}|\pm1\rangle
=
\frac{A_{\rm xz}}{\sqrt2}I_\mathrm{z} .
\end{equation}

The second-order correction is therefore

\begin{equation}
\propto
\frac{A_{\rm xz}^2}{2}
I_\mathrm{z}^2 .
\end{equation}

For a spin-$1/2$ nucleus,

\begin{equation}
I_\mathrm{z}^2=\frac14 ,
\end{equation}

so this term produces only a scalar energy correction and does not generate a transverse nuclear operator.

The mixed electronic-Zeeman/hyperfine contribution gives

\begin{equation}
Z_\pm H_\pm^\dagger
\propto
B_\pm I_\mathrm{z} ,
\end{equation}

which only modifies the longitudinal nuclear Hamiltonian.

Therefore,

\begin{equation}
\boxed{
A_{\rm xz},A_{\rm yz}
\text{ do not contribute to the transverse hyperfine-enhanced nuclear Rabi
coupling.}
}
\end{equation}

\subsection*{Transverse Flip-Flop Terms}

The terms

\begin{equation}
A_{\rm xx}S_{\rm x} I_{\rm x}+A_{\rm yy}S_{\rm y}I_{\rm y}
\end{equation}

contain

\begin{equation}
S_{\rm x}I_{\rm x}+S_{\rm y}I_{\rm y}
=
\frac12(S_+I_-+S_-I_+),
\end{equation}

which directly couples electronic and nuclear spin flips.

The corresponding matrix elements are

\begin{align}
\langle0|H_{\rm HF}|+1\rangle
&=
\frac{A_\perp}{\sqrt2}I_- ,
\\
\langle0|H_{\rm HF}|-1\rangle
&=
\frac{A_\perp}{\sqrt2}I_+ ,
\end{align}

where

\begin{equation}
A_\perp^2=A_{\rm xx}^2+A_{\rm yy}^2 .
\end{equation}

These terms generate the effective transverse nuclear field

\begin{equation}
\nu_\perp
=
-\frac{2\gamma_{\rm e} B_\perp A_\perp D}
{D^2-(\gamma_{\rm e}B_{\rm z})^2},
\end{equation}

which produces the observed enhanced nuclear oscillations.

\clearpage
\section{Contrast of nuclear oscillation}
Equation 6 in the main text can be described as an effective, factorized description of the measured signal, in which the observed contrast is expressed as the product of two independently determined contributions: the electron spin readout contrast and the nuclear precession amplitude.

This form follows from the assumption that the nuclear spin dynamics and the optical readout process remain approximately separable, such that the fluorescence signal depends linearly on the mapped electron spin population. 
\subsection{Derivation for the electron spin}
We define the effective electronic field components as
\begin{equation}
\Omega_\mathrm{z}^{(\mathrm{e})} = \gamma_\mathrm{e} B_\mathrm{z} - A_\parallel, 
\qquad
\Omega_\mathrm{x}^{(\mathrm{e})} = \gamma_\mathrm{e} B_\perp,
\end{equation}
such that the electronic transition frequency is
\begin{equation}
f_\mathrm{e} = \sqrt{(\Omega_\mathrm{z}^{(\mathrm{e})})^2 + (\Omega_\mathrm{x}^{(\mathrm{e})})^2}.
\end{equation}

Introducing the effective electronic angle
\begin{equation}
\theta_\mathrm{e} = \tan^{-1}\!\left(\frac{\Omega_\mathrm{x}^{(\mathrm{e})}}{\Omega_\mathrm{z}^{(\mathrm{e})}}\right)
= \tan^{-1}\!\left(\frac{\gamma_\mathrm{e} B_\perp}{\gamma_\mathrm{e} B_\mathrm{z} - A_\parallel}\right),
\end{equation}
we obtain
\begin{equation}
\sin\theta_\mathrm{e} = \frac{\Omega_\mathrm{x}^{(\mathrm{e})}}{f_\mathrm{e}}, 
\qquad
\cos\theta_\mathrm{e} = \frac{\Omega_\mathrm{z}^{(\mathrm{e})}}{f_\mathrm{e}}.
\end{equation}

The electronic contrast can therefore be written as
\begin{equation}
C_\mathrm{e} \sim \left(\frac{\gamma_\mathrm{e} B_\mathrm{z} - A_\parallel}{f_\mathrm{e}}\right)^2
= \cos^2\theta_\mathrm{e}.
\end{equation}

\subsection{Derivation for the nuclear spin}
Similarly, defining the effective nuclear field components
\begin{equation}
\Omega_\mathrm{z} = \gamma_\mathrm{I} B_\mathrm{z} + \nu_\mathrm{z}, 
\qquad
\Omega_\mathrm{x} = \gamma_\mathrm{I} B_\perp + \nu_\perp,
\end{equation}
and the effective nuclear angle
\begin{equation}
\vartheta_{\mathrm{eff}} = \tan^{-1}\!\left(\frac{\Omega_\mathrm{x}}{\Omega_\mathrm{z}}\right),
\end{equation}
the nuclear contrast becomes
\begin{equation}
C_\mathrm{n} = \left(\frac{\Omega_\mathrm{x}}{f_{\mathrm{nucl}}}\right)^2
= \sin^2\vartheta_{\mathrm{eff}},
\end{equation}
with
\begin{equation}
f_{\mathrm{nucl}} = \sqrt{\Omega_\mathrm{z}^2 + \Omega_\mathrm{x}^2}.
\end{equation}

Thus, the total contrast can be written compactly as
\begin{equation}
C \sim \cos^2\theta_\mathrm{e} \, \sin^2\vartheta_{\mathrm{eff}}.
\end{equation}
\clearpage

\section{Magnetic field optimization}
In addition to the method described in the main part for aligning the magnet using nuclear spin precessions, there are three other methods, which are described below.
\subsection{Optimization via count rate}
The simplest and fastest method of aligning the magnetic field is by optimizing the count rate of a single PL6 center. For PL6 centers, the detected count rate is maximum when the magnetic field is aligned due to the inter system crossing rates of the Spin 1 system. The misalignment results in spin mixing, which partially pumps the system to the $m_\mathrm{s} = 1$ or $m_\mathrm{s} = -1$ state, causing the count rate to decrease. For the measurements, the lateral position of the magnet is swept and the count rate is measured for a fixed magnetic field.
This is shown in Supplementary Figure \ref{Fig_S4_Magnet}a for an applied external magnetic field. By fitting a 2D Gaussian function, the maximum count rate and thus the position with the best alignment can be found.

\begin{figure}[ht]
\includegraphics[width=\linewidth]{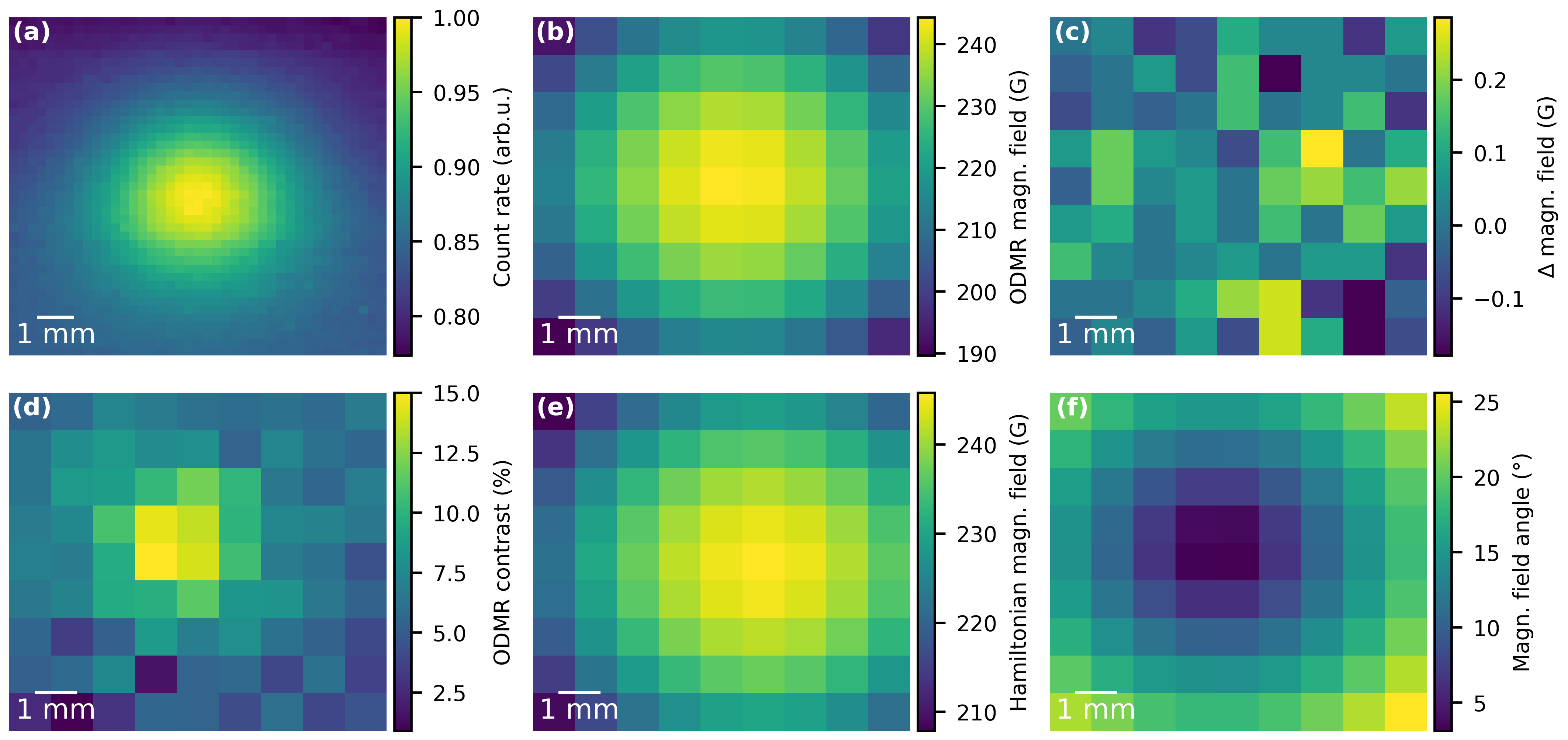}
\centering
\caption[]{\textbf{Magnetic field optimization.} All measurements were done by sweeping the lateral position of the magnet in a fixed height above the sample. \textbf{(a)} Count rate of a single PL6 center in dependence of the external magnetic field. \textbf{(b)} Measured magnetic field via the $\ket{-1}$ and $\ket{+1}$ ODMR transitions. \textbf{(c)} Reproducibility study of the measurement in \textbf{(b)} by remeasuring the magnetic field and determining the differences compared to the first measurement to show a repetition and exact alignment of the magnet is given. \textbf{(d)} Contrast of the $\ket{+1}$ transition in dependence of the magnetic field position. \textbf{(e) \& (f)} Calculated external magnetic field strength and angle via the Hamiltonian. 
% If figure twocolumn: Put * behind figure
}
\label{Fig_S4_Magnet}
\end{figure}

\subsection{Optimization via ODMR measurements}
Another way to determine alignment is with ODMR measurements. When the magnetic field is aligned, the ODMR splitting between the two transitions is maximal, while it is reduced for deviations in alignment.

As with the count rate measurements, the lateral position of the magnet is swept, and for a single PL6 center, the ODMR transition $m\mathrm{_s} = 0 \rightarrow m\mathrm{_s} = \pm 1$ is measured and fitted with a 2D Gaussian function to find the ideal magnet position. From the difference between the two ODMR frequencies one can calculate the magnetic field strength  which is depicted in Supplementary Figure \ref{Fig_S4_Magnet}b. 
Furthermore, it is important to determine whether a reproducible magnet position can be achieved to enable precise alignment. A repetition of the experiment in Supplementary Figure \ref{Fig_S4_Magnet}b and subsequent comparison are shown in Supplementary Figure \ref{Fig_S4_Magnet}c. It shows that only minimal differences in the magnetic field arise, which can be explained by the fitting uncertainty of the ODMR measurements. In addition, a clear dependence of the contrast in the ODMR measurement can be seen, which is shown in Supplementary Figure \ref{Fig_S4_Magnet}d.

\subsection{Optimization via Hamiltonian}
With the ODMR frequencies $\nu_1$ and $\nu_2$ already determined, the alignment can also be done using the Hamiltonian. This was already shown by Balasubramanian et al. in 2008 \cite{Balasubramanian2008NanoscaleConditions}. Here, only an abbreviated derivation is discussed.

In summary, from the reduced Hamiltonian to the electron spin from equation 1 in the main part, the magnetic field is characterized by the absolute strength $B$ and the two polar and azimuthal angles $\Theta$ and $\varphi$ by transforming the coordinates into polar coordinates. The position of the spin levels can be determined by solving the following equation for $x$
% \begin{equation}
%     H = \mu_B g \boldsymbol{B} \cdot \boldsymbol{S} + D (S_z^2-S(S+1/3)) + E(S_x^2-S_y^2)
% \end{equation}

\begin{equation}
    x^3 - \Big(\frac{D^2}{3} + E^2 + \beta^2\Big)x - \frac{\beta^2}{2}\Big(D\cos(2\Theta) + 2E\cos(2\phi)\sin^2(\Theta)\Big) - \frac{D}{6}\Big(4E^2+\beta^2\Big) + \frac{2D^3}{27} = 0
\end{equation}
with $\beta = \mu_\mathrm{B}gB$. The relation
\begin{equation}
    \Delta = D\cos(2\Theta) + 2E\cos(2\phi)\sin^2(\Theta)
\end{equation}
leads to 
\begin{equation}
    \beta^2 = \frac{1}{3}\Big(\nu^2_1+\nu_2^2-\nu_1\nu_2-D^2\Big)-E^2
\end{equation}
and 
\begin{equation}
    \Delta = \frac{7D^3+2(\nu_1+\nu_2)(2(\nu_1^2+\nu_2^2)-5\nu_1\nu_2-9E^2)-3D(\nu^2_1+\nu_2^2-\nu_1\nu_2+9E^2)}{9(\nu^2_1+\nu_2^2-\nu_1\nu_2-D^2-3E^2)} \quad \xrightarrow{D\gg E} \quad\Delta \approx D\cos(2\Theta)
\end{equation}
for solutions for the magnetic field strength and the tilt angle and are illustrated in Supplementary Figures \ref{Fig_S4_Magnet}e \& \ref{Fig_S4_Magnet}f.

\clearpage
\section{Tomography}
A more detailed description of the tomographies of electron spin, nuclear spin and the coupled two-qubit system shown in Figure 5 in the main part follows here.
\subsection{Electron spin}
For the simple one-qubit case, electron spin tomography can be determined using Rabi oscillations. The system is initialized to the $m_\mathrm{s}=0$ state, in this case the transition $m_\mathrm{s}=0 \rightarrow m_\mathrm{s}=-1 $ is driven by a resonant microwave pulse, and finally, before the readout, one of three possible pulses is taken to measure the individual bases. No additional pulse is taken for the $Z$ basis; for $X$ and $Y$, a $\nicefrac{\pi}{2}$ pulse is applied in the $X$- or $Y$-direction. This is achieved using the 90 $^\circ$ phase shifter.
The corresponding pulse sequence is shown schematically in Supplementary Figure \ref{Fig_S5_Electron}a. \\

\begin{figure}[ht]
\includegraphics[width=0.8\linewidth]{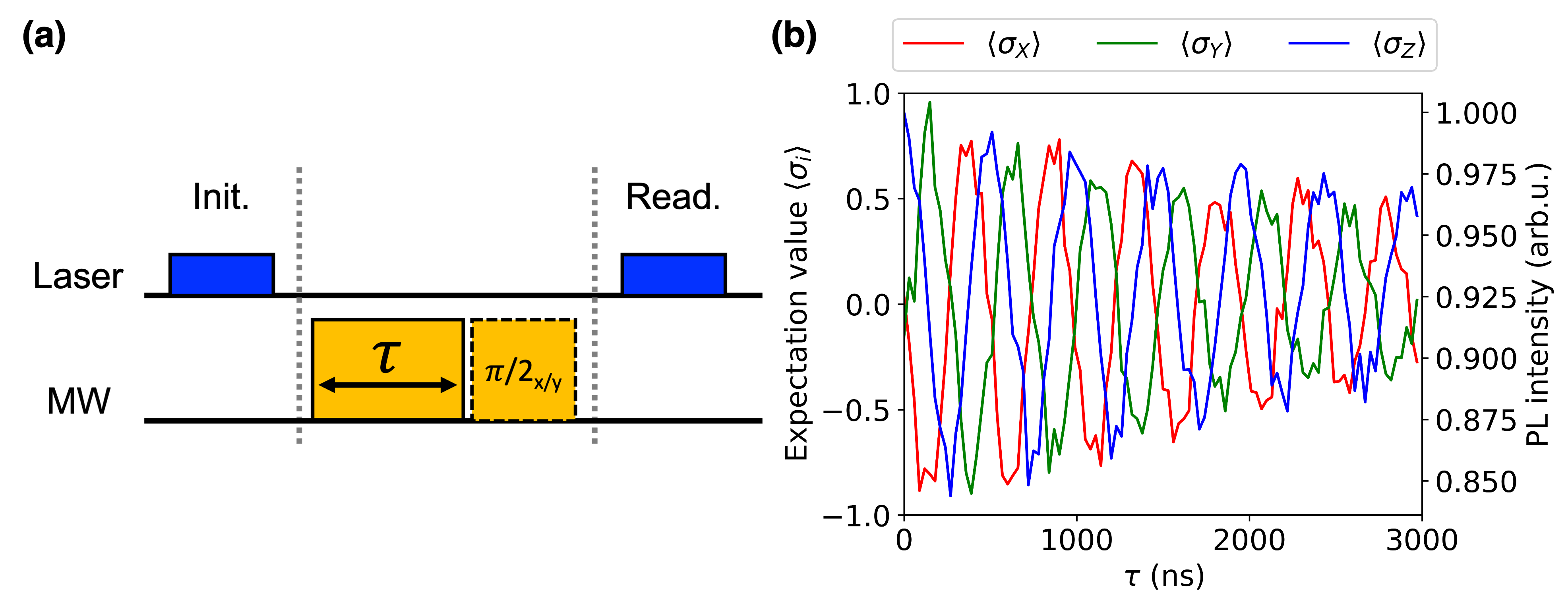}
\centering
\caption[]{\textbf{Electron spin tomography.} \textbf{(a)} Measurement sequence for the electron spin tomography. For the $Z$ basis, no additional pulse is applied whereas for the $X$ and $Y$ basis, a $\nicefrac{\pi}{2}$ pulse in the corresponding direction is applied. \textbf{(b)} Experimental result for the electron spin tomography. The determined expectation values can be seen in Table \ref{tab:Tomography_electron}.
% If figure twocolumn: Put * behind figure
}
\label{Fig_S5_Electron}
\end{figure}

Experimentally, the following three expectation values for $X$, $Y$, and $Z$ are obtained, which are shown in Supplementary Table \ref{tab:Tomography_electron}. The absolute measured Rabi contrast is normalized to obtain an expectation value between -1 and 1. This is shown in Supplementary Figure \ref{Fig_S5_Electron}b. The first data point was used for the tomography because, as can be seen experimentally, the Rabi oscillations show exponential decay, and thus a false result would otherwise be formed.

\begin{table}[ht]
    \caption{\textbf{Electron spin tomography.} Experimentally determined expectation values for the electron spin.}
    \centering
    \begin{tabular}{c|c}
        $S_\mathrm{i}$ & $\langle{S_\mathrm{i}}\rangle$ \\\hline
        $Z$ & 0.9353\\
        $X$ & 0.0394\\
        $Y$ & -0.1555\\
    \end{tabular}
    \label{tab:Tomography_electron}
\end{table}
Using these three expectation values, the density matrix for a one-qubit system can be determined. This works as follows:
\begin{equation}
\rho = \frac{1}{2} \Big(I + \sum_{\mathrm{i}} \langle S_\mathrm{i}\rangle \sigma_\mathrm{i} \Big)
\end{equation}
where $I$ is the identity matrix and $\sigma_\mathrm{i}$ are the Pauli matrices.

\begin{equation}
    I = \begin{pmatrix}
        1 & 0 \\
        0 & 1 \\
    \end{pmatrix},\quad
    X = \begin{pmatrix}
        0 & 1 \\
        1 & 0 \\
        \end{pmatrix}, \quad
        Y = \begin{pmatrix}
        0 & -i \\
        i & 0 \\
    \end{pmatrix},\quad
    Z = \begin{pmatrix}
        1 & 0 \\
        0 & -1 \\
    \end{pmatrix}
\end{equation}
Using the experimental parameters, the density matrix can be reconstructed, and to determine the fidelity, the experimentally determined state is compared to the ideal state $\ket{\Psi^+}$ with
\begin{equation}
\ket{\Psi^+} = \begin{pmatrix}
1 & 0 \\
0 & 0 \\
\end{pmatrix}
\end{equation}
and the fidelity $F$ is determined as follows
 \begin{equation}
    F = \mathrm{Tr}\Big( \sqrt{\sqrt{\rho} \rho'\sqrt{\rho}}\Big)
\end{equation}
where $\rho$ is the experimentally determined density matrix and $\rho'$ the Bell state matrix.

\subsection{Nuclear spin}
For nuclear spin, tomography works in theory in exactly the same way as for electron spin. However, the nuclear spin cannot be read directly; therefore, as in all measurements, it is projected onto the electron spin to be read out.
\begin{figure}[ht]
\includegraphics[width=0.8\linewidth]{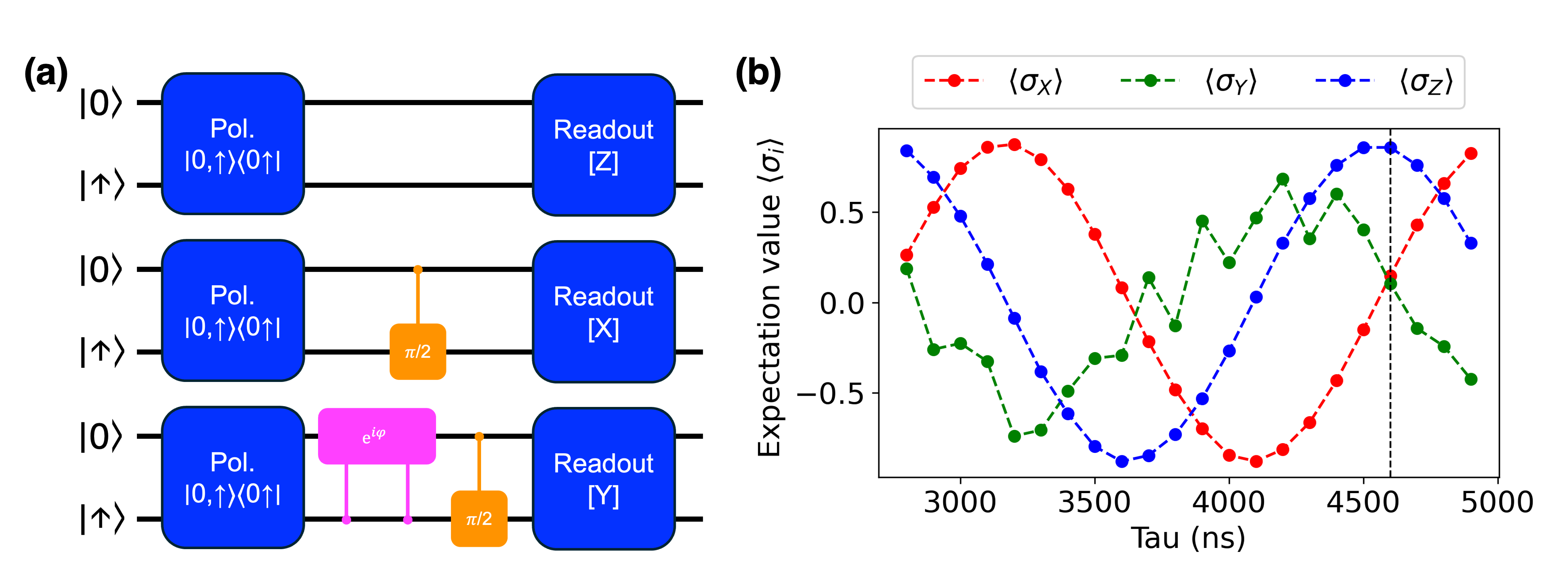}
\centering
\caption[]{\textbf{Nuclear spin tomography.} \textbf{(a)} Measurement sequence for the nuclear spin tomography. For the $Z$ basis, no additional pulse is applied. For the $X$ basis, a waiting time of $\nicefrac{\pi}{2}$ is applied. For the $Y$ basis, additionally a phase is acquired using a MW $2\pi$ pulse with a subsequent waiting time to achieve the 90 $^\circ $ rotation. \textbf{(b)} Experimental result for the nuclear spin tomography. The  determined expectation values can be seen in Table \ref{tab:Tomography_nuclear}.

% If figure twocolumn: Put * behind figure
}
\label{Fig_S5_Nuclear}
\end{figure}

For $Z$ and $X$, tomography works similarly to electron spin. The measured value is used directly for the expectation value in $Z$, while a $\nicefrac{\pi}{2}$ pulse in the x-direction is used for the expectation value in $X$. In this case, the $\nicefrac{\pi}{2}$ pulse corresponds exactly to one waiting time of the nuclear spin precession by $\nicefrac{\pi}{2}$. For the y-direction, it is not so trivial, since the method we described cannot control the precession in $Y$. However, this can be remedied by using the single pulse in $X$-direction and additionally adding a phase of the nuclear spin. Illustratively, the nuclear spin precesses when the electron spin is in the $m_\mathrm{s}=0$ state. If a $2\pi$ pulse is applied, the nuclear spin acquires a phase in the state $m_\mathrm{s}=1$, relatively to the $m_\mathrm{s}=0$ state. In our case, we apply two "fast" individual $\pi$ pulses with a waiting time of $\tau_{\mathrm{wait}} \approx 10$ ns between them. This value was chosen as a shorter waiting time would not be experimentally implementable and for longer waiting times the contrast of the nuclear precession would vanish due to decoherence. 
The schematic representation of the measurement of the three bases is shown in Supplementary Figure \ref{Fig_S5_Nuclear}a. The experimental results are as follows: the expected values for the nuclear spin in its three bases oscillate with its precession frequency. For the tomography, the value at which the desired tomography state has a maximum $Z$ value and minimal $X$ and $Y$ values is taken. In this example, this is the case at $\tau = 4600$ ns and is plotted in Supplementary Figure \ref{Fig_S5_Nuclear}b. 

\begin{table}[ht]
    \caption{\textbf{Nuclear spin tomography.} Experimentally determined expectation values for the nuclear spin.}
    \centering
    \begin{tabular}{c|c}
        $I_\mathrm{i}$ & $\langle{I_\mathrm{i}}\rangle$ \\\hline
        $Z$ & 0.8574\\
        $X$ & 0.1475\\
        $Y$ & 0.1050\\
    \end{tabular}
    \label{tab:Tomography_nuclear}
\end{table}

The exact values for the tomography are shown in Supplementary Table \ref{tab:Tomography_nuclear}. Using these values, as for the electron spin, the density matrix can be reconstructed and the fidelity determined, as already shown in the main section.

\subsection{Two-qubit tomography}
For the two-qubit tomography, the following change in the density matrix results compared to the one-qubit tomography
\begin{equation}
\rho = \frac{1}{4} \Big(I \otimes I + \sum_{\mathrm{i},\mathrm{j}} \langle S_\mathrm{i}I_\mathrm{j}\rangle S_\mathrm{i} \otimes I_\mathrm{j}\Big)
\end{equation}
and the following Bell state, which should be created.
\begin{equation}
    |\Psi \rangle = \frac{1}{2} 
    \begin{pmatrix}
        1 & 0 & 0 & 1\\
        0 & 0 & 0 & 0\\
        0 & 0 & 0 & 0\\
        1 & 0 & 0 & 1\\
    \end{pmatrix}    
\end{equation}

The entangled state is achieved, as shown in Figure 5c in the main part, by a global $\pi$ pulse on the electron spin and half a Larmor precession. Global here means that the pulse is on both ODMR transitions, i.e. both spin up and spin down for the nuclear spin.
For tomography, in contrast to the one-qubit case, not 3, but 9 measurements are required to measure all bases. This is achieved by combining the 3 bases for the nuclear spin with the three for the electron spin from Supplementary Figures \ref{Fig_S5_Electron}a and \ref{Fig_S5_Nuclear}a.
The nuclear spin is then projected onto the electron spin for readout. In this case, timing is extremely critical, especially since the microwave pulses are sometimes very short (30 ns for the $Y$ rotation of the nuclear spin). This ensured that, for example, the waiting time for a nuclear spin precession was from the center of the microwave pulse to the center of the next pulse in order to achieve an exact rotation.

In addition, unlike the other measurements, a time-dependent measurement where one waiting time or pulse length is swept is not possible, as otherwise the state cannot be maintained and a mixed state would settle. Therefore, each measurement was performed with a fixed time and the expected value was determined using reference measurements.
For the measurements, two reference measurements were taken, one with minimum and one with maximum contrast, to which all subsequent measurements were normalized.

\begin{figure}[ht]
\includegraphics[width=0.4\linewidth]{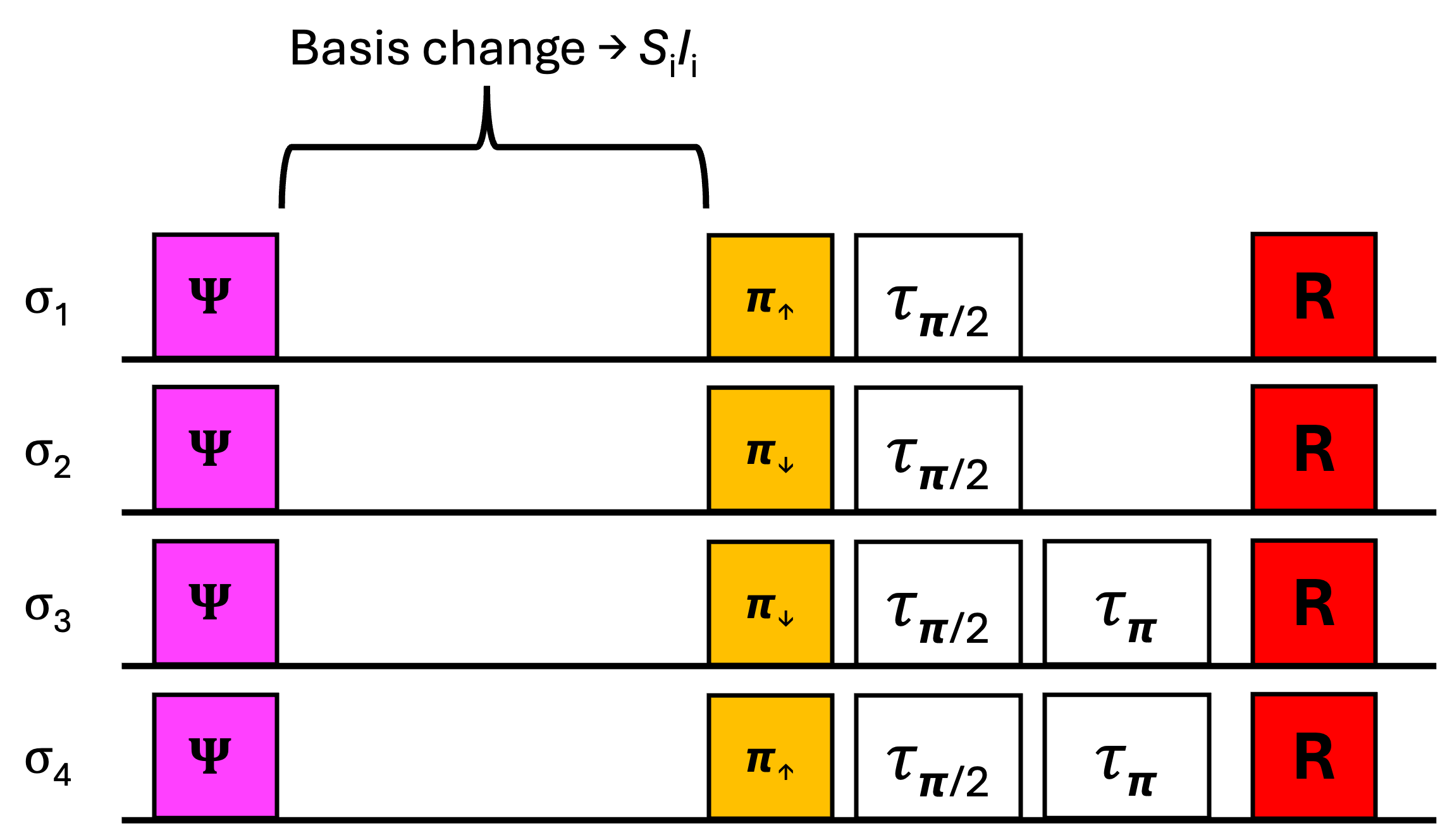}
\centering
\caption[]{\textbf{Tomography readout.} Schematic representation of all four different readout possibilities to measure the expectation value needed for the two qubit tomography. The fourth readout was not performed due to the constraint that the sum of all $\sigma_{\mathrm{i}{_\mathrm{i}}}$ equals to one.
% If figure twocolumn: Put * behind figure
}
\label{Fig_S5_Readout}
\end{figure}
For readout, the system was then brought into one of the two electronic states, $m_\mathrm{s}=0$ or $m_\mathrm{s}=1$, using pulse combinations. This is shown schematically in Supplementary Figure \ref{Fig_S5_Readout}.
With these three measurements, the coefficients $\sigma_{\mathrm{i}{_1}}$, $\sigma_{\mathrm{i}{_2}}$, and $\sigma_{\mathrm{i}{_3}}$ can be determined, which are necessary for calculating the expectation values. The required fourth value can be determined using the relation $\sum_{\mathrm{i}} \sigma_{\mathrm{i}{_1}} = 2$. The 2 arises from the fact that, ideally, two measurements show maximum contrast, i.e., $\sigma_{\mathrm{i}{_\mathrm{i}}} = 1$, while two have minimum contrast, i.e., $\sigma_{\mathrm{i}{_\mathrm{i}}} = 0$, and thus the sum of these is 2.
By halving the values, this can be normalized to 1 and using the relation $\langle \sigma_{\mathrm{i}} \rangle = \sigma_{\mathrm{i}{_1}} - \sigma_{\mathrm{i}{_2}} - \sigma_{\mathrm{i}{_3}} + \sigma_{\mathrm{i}{_4}}$ the expectation value can be determined.

\begin{table}[ht]
    \caption{\textbf{Two-qubit tomography.} Experimental results for the two-qubit tomography with all three experimentally determined parameters $\sigma_{\mathrm{i}{_\mathrm{i}}}$ used for calculating the expectation value $\langle S_\mathrm{i}I_\mathrm{j}\rangle $ for each basis.}
    \centering
    \begin{tabular}{c|c||c|c|c||c}
        $S_\mathrm{i}$ & $I_\mathrm{j}$ & $\sigma_{\mathrm{ij}{_1}}$ & $\sigma_{\mathrm{ij}{_2}}$ & $\sigma_{\mathrm{ij}{_3}}$ & $\langle S_\mathrm{i}I_\mathrm{j}\rangle $ \\\hline\hline
        $Z$ & $Z$ &0.982 &0.018 & 0.058&  0.924\\
        $X$ & $X$ & 0.891& 0.140& 0.009& 0.851\\
        $Y$ & $Y$ & -0.060& 0.931& 0.852& -0.783\\
        $X$ & $Y$ & 0.433& 0.385& 0.471& 0.144\\
        $X$ & $Z$ & 0.347& 0.496& 0.541& -0.037\\
        $Y$ & $X$ & 0.392& 0.487& 0.477& 0.036\\
        $Y$ & $Z$ & 0.486& 0.526& 0.473& 0.001\\
        $Z$ & $X$ & 0.434& 0.495& 0.567& -0.062\\
        $Z$ & $Y$ & 0.551& 0.460& 0.443& 0.097\\
    \end{tabular}
    \label{tab:Tomography}
\end{table}

The results of the measurements are shown in Supplementary Table \ref{tab:Tomography}.

Analogous to the other two tomographies of the one qubit systems, the density matrix is reconstructed using the expected values and compared with the ideal Bell state, which was already described in the main text.

\bibliography{Supplementary}
\newpage